\documentclass[conference,10pt]{IEEEtran}
\IEEEoverridecommandlockouts
\usepackage{amsmath,amssymb,amsfonts,mathtools,nccmath,mathrsfs,amsthm}
\usepackage{algorithmic}
\usepackage{algorithm}
\usepackage{graphicx}
\usepackage{array}
\usepackage{subfigure}
\usepackage{textcomp}
\usepackage{xcolor}
\usepackage{soul}
\usepackage{cite}
\def\BibTeX{{\rm B\kern-.05em{\sc i\kern-.025em b}\kern-.08em T\kern-.1667em\lower.7ex\hbox{E}\kern-.125emX}}
\usepackage{hyperref}
\usepackage[font=small,skip=2pt]{caption}

\usepackage{tikz}
\usepackage{pgfplots}
\usetikzlibrary{decorations.pathreplacing}
\usepackage{interval}
\usetikzlibrary{arrows,shapes,backgrounds,plotmarks,positioning}

\pgfplotsset{compat=newest} % move axis labels close to the tick label automatically
\pgfplotsset{plot coordinates/math parser=false}
\newlength\figureheight
\newlength\figurewidth

\newtheorem{definition}{Definition}

\newtheorem{proposition}{Proposition}

\newcommand{\X}{\mathcal{X}}
\newcommand{\Y}{\mathcal{Y}}
\newcommand{\Sen}{\mathcal{S}}
\newcommand{\PS}{P_{S}}
\newcommand{\PX}{P_{X}}
\newcommand{\PY}{P_{Y}}
\newcommand{\PSX}{P_{S,X}}
\newcommand{\PSgX}{P_{S|X}}
\newcommand{\PXgS}{P_{X|S}}

\newcommand{\PYgX}{P_{Y|X}}
\newcommand{\PSY}{P_{S,Y}}
\newcommand{\PSgY}{P_{S|Y}}
\newcommand{\PYgS}{P_{Y|S}}

\newcommand{\E}{\mathbb{E}}
\newcommand{\Set}[1]{\{#1\}}

\newcommand{\RealP}{\mathbb{R}_+}
\newcommand{\argmin}{\arg\!\min}
\newcommand{\argmax}{\arg\!\max}
\newcommand{\eps}{\varepsilon}

\newcommand{\lif}{\ell(s,x)}
\newcommand{\lglif}{i(s,x)}

\newcommand{\epsu}{\varepsilon_{u}}
\newcommand{\epsl}{\varepsilon_{l}}

\begin{document}

\title{Asymmetric Local Information Privacy and the Watchdog Mechanism\\
\thanks{The work of P.~Sadeghi  and M.~A.~Zarrabian is supported by the ARC Future Fellowship FT190100429 and partly by the Data61 CRP: IT-PPUB.}
}
%1\textsuperscript{st}
\author{\IEEEauthorblockN{Mohammad Amin Zarrabian}
\IEEEauthorblockA{\textit{College of Engineering}\\ \textit{and Computer Science,} \\
\textit{Australian National University,}\\
Canberra, Australia.\\
mohammad.zarrabian@anu.edu.au}
\and
%2\textsuperscript{nd}
\IEEEauthorblockN{Ni Ding}
\IEEEauthorblockA{\textit{School of Computing}\\ \textit{and Information Systems,} \\
\textit{University of Melbourne,}\\
Melbourne, Australia. \\
ni.ding@unimelb.edu.au}
\and
%3\textsuperscript{rd}
\IEEEauthorblockN{Parastoo Sadeghi}
\IEEEauthorblockA{\textit{School of Engineering}\\ \textit{and Information Technology,} \\
\textit{University of New South Wales,}\\
Canberra, Australia. \\
p.sadeghi@unsw.edu.au}
}

\maketitle
%%%%%%%%%%%%%%%%%%%%%%%%%%%%%%%%%%%%%%%%%%%%%%%%%%%%%%%%%%%%%%%%%%%%%%%%%%%%%%%%%%%%%%%%%%%%%%%%%%%%%%%%%%%%%%%%%%%%%%%%%%%%%%%%%%%%%%%%%%%%%%%%%%%%%%%%%%%%%%%%%%%%%%%%%%%%%%%%%%
\begin{abstract}
	This paper proposes a novel watchdog privatization scheme by generalizing local information privacy (LIP) to enhance data utility.
	To protect the sensitive features $S$ correlated with some useful data $X$, LIP restricts the lift, the ratio of the posterior belief to the prior on $S$ after and before accessing $X$.
	For each $x$, both maximum and minimum lift over sensitive features are measures of the privacy risk of publishing this symbol and should be restricted for the privacy-preserving purpose. 
	Previous works enforce the same bound for both max-lift and min-lift.
	However, empirical observations show that the min-lift is usually much smaller than the max-lift.
	In this work, we generalize the LIP definition to consider the unequal values of max and min lift, i.e., considering different bounds for max-lift and min-lift. 
	This new definition is applied to the watchdog privacy mechanism.
	We demonstrate that the utility is enhanced under a given privacy constraint on local differential privacy.
	At the same time, the resulting max-lift is lower and, therefore, tightly restricts other privacy leakages, e.g., mutual information, maximal leakage, and $\alpha$-leakage.
\end{abstract}


%%%%%%%%%%%%%%%%%%%%%%%%%%%%%%%%%%%%%%%%%%%%%%%%%%%%%%%%%%%%%%%%%%%%%%%%%%%%%%%%%%%%%%%%%%%%%%%%%%%%%%%%%%%%%%%%%%%%%%%%%%%%%%%%%%%%%%%%%%%%%%%%%%%%%%%%%%%%%%%%%%%%%%%%%%%%%%%%%%
\section{Introduction}
	Today, many businesses and government agencies gather and share massive amounts of data to achieve economic and social benefits through the widespread advancement of communication systems and machine learning algorithms.
	This phenomenon raises a growing concern about the privacy of individual users that could be at risk by inferring confidential information from datasets explicitly or implicitly. 
	This situation motivates research to design privacy-preserving mechanisms that, besides protecting confidential features, also provide a satisfactory data utility.
	
	The information-theoretic (IT) paradigm measures privacy as information leakage about private data $S$ when correlated useful data $X$ is accessed. 
	In this regard, the lift is a symbol-wise metric of the privacy \cite{2012_privacy_statisticalinference,2019Watchdog}, which determines the adversary's knowledge gain via measuring the multiplicative gain of posterior belief $\PSgX(s|x)$ compared to the prior $\PS(s)$, and is given by:
	\begin{equation}\label{eq:lift}
		\lif=\frac{\PSgX(s|x)}{\PS(s)}.
	\end{equation}
	The logarithm of the lift is called \textit{log-lift} and denoted by $\lglif=\log\lif$.
	Lift and log-lift provide strong notions of IT privacy in \textit{local information privacy} (LIP) \cite{2012_privacy_statisticalinference,2019Watchdog,2020SadeghiITW,2021linearreduct,2018ContextdataAggre,2019Infoprivacybounded,2019LIPLaplcian,2020LIPDATAAGG,2021Contextaware}, where a sanitized version of $X$ is published to restrict $|\lglif|$ below a given threshold $\eps$ known as \textit{privacy budget}.
	$\eps$-LIP upper bounds IT measures including the mutual information (MI) \cite{2012_privacy_statisticalinference,2014_Info_Bottl_PF}, maximal leakage \cite{2016IssaMaxL}, $\alpha$-leakage, $\alpha$-lift \cite{2021alphading}, and Sibson MI \cite{2015VerduAlphaMI,2018_tunable_measureinformationleakage,2021alphading}. 
	It also provides a $2\eps$ upper bound on the local differential privacy (LDP) \cite{2011learnprivately,2013LDPMiniMax}.

	To attain LIP, \cite{2019Watchdog} has proposed a watchdog privacy mechanism in which  the alphabet of $X$, denoted by $\X$,  is bi-partitioned into subsets of low-risk $\X_L=\Set{x\in\X:\max_{s}|i(s,x)|\leq\eps}$ and high-risk, $\X_H=\X\setminus\X_L$ symbols. 
	Then, the low-risk subset is published without alteration and only high-risk symbols are perturbed via a randomization $r_{Y|X}(y|x)$.
	It has been proved in \cite{2020SadeghiITW} that any $X$-invariant $r_{Y|X}(y|x)$, e.g., merging all symbols in $\X_H$ minimizes the privacy leakage in $\X_H$.
	Using $X$-invariant randomization to achieve high level privacy protection could result in low utility. 

	This paper proposes a method to enhance the utility based on the finer-grained properties of the log-lift.   
	From the intuition behind \eqref{eq:lift}, one can see that a large lift value means that observing $x$ increases the prior belief about the associated sensitive feature $s$, $\left(\lif>1 \Rightarrow \PSgX(s,x) > \PS(s)\right)$.
	In contrast, a lift value less than one means releasing $x$ decreases the posterior belief.
	Consequently, both large and small lift values are measures of higher leakage about sensitive features.
	To see this clearly, we generalize LIP measures and define the following quantities:
	\begin{equation}\label{eq:max-min-lift}
		\nu(x)=\min_{s}\lglif, \quad \xi(x)= \max_{s}\lglif.
	\end{equation}
	%%%%%
	
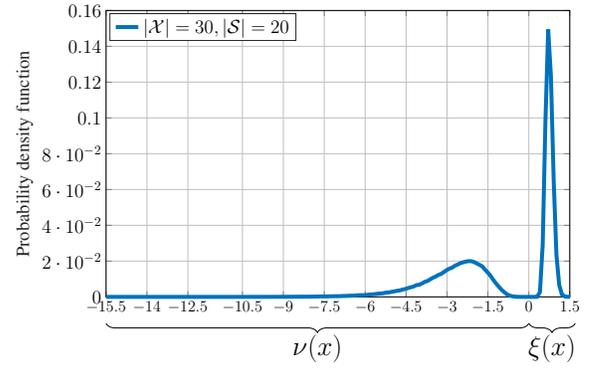
\begin{figure}[]
		\centering
		\scalebox{0.5}{\definecolor{mycolor1}{rgb}{0.00000,0.44700,0.74100}%
\definecolor{mycolor2}{rgb}{0.85000,0.32500,0.09800}%
\begin{tikzpicture}

\begin{axis}[%
width=4.854in,
height=3in,
scale only axis,
xmin=-15.5,
xmax=1.5,
xtick={-15.5,-14,-12.5,-10.5,-9,-7.5,-6,-4.5,-3,-1.5,0,1.5},
every x tick label/.append style={font=\color{darkgray!60!black},font=\large},
xlabel style={font=\color{white!15!black}, yshift=-20pt},
ymin=0,
ymax=0.16,
every y tick label/.append style={font=\color{darkgray!60!black},font=\Large},
ylabel style={font=\color{white!15!black}},
ylabel={\Large Probability density function},
axis background/.style={fill=white},
clip=false,
xmajorgrids,
ymajorgrids,
legend style={at={(0.41,0.99)},draw=darkgray!60!black,fill=white,legend cell align=left, font=\Large},
]
\addplot [color=mycolor1, line width=3.0pt]
  table[row sep=crcr]{%
-15.4965631641459	1.66666666666667e-06\\
-15.3959055123505	1.66666666666667e-06\\
-15.2952478605551	0\\
-15.1945902087596	0\\
-15.0939325569642	0\\
-14.9932749051687	0\\
-14.8926172533733	0\\
-14.7919596015778	0\\
-14.6913019497824	0\\
-14.5906442979869	0\\
-14.4899866461915	0\\
-14.389328994396	0\\
-14.2886713426006	0\\
-14.1880136908052	0\\
-14.0873560390097	1.66666666666667e-06\\
-13.9866983872143	0\\
-13.8860407354188	0\\
-13.7853830836234	0\\
-13.6847254318279	0\\
-13.5840677800325	1.66666666666667e-06\\
-13.483410128237	3.33333333333333e-06\\
-13.3827524764416	1.66666666666667e-06\\
-13.2820948246461	0\\
-13.1814371728507	0\\
-13.0807795210553	1.66666666666667e-06\\
-12.9801218692598	0\\
-12.8794642174644	0\\
-12.7788065656689	1.66666666666667e-06\\
-12.6781489138735	1.66666666666667e-06\\
-12.577491262078	3.33333333333333e-06\\
-12.4768336102826	0\\
-12.3761759584871	0\\
-12.2755183066917	0\\
-12.1748606548962	0\\
-12.0742030031008	1e-05\\
-11.9735453513054	5e-06\\
-11.8728876995099	6.66666666666667e-06\\
-11.7722300477145	1.66666666666667e-06\\
-11.671572395919	0\\
-11.5709147441236	5e-06\\
-11.4702570923281	1.66666666666667e-06\\
-11.3695994405327	3.33333333333333e-06\\
-11.2689417887372	6.66666666666667e-06\\
-11.1682841369418	3.33333333333333e-06\\
-11.0676264851463	5e-06\\
-10.9669688333509	5e-06\\
-10.8663111815555	0\\
-10.76565352976	6.66666666666667e-06\\
-10.6649958779646	1e-05\\
-10.5643382261691	1e-05\\
-10.4636805743737	1.16666666666667e-05\\
-10.3630229225782	1.66666666666667e-05\\
-10.2623652707828	1.5e-05\\
-10.1617076189873	3.16666666666667e-05\\
-10.0610499671919	2.66666666666667e-05\\
-9.96039231539645	1.66666666666667e-05\\
-9.859734663601	1.83333333333333e-05\\
-9.75907701180556	2.16666666666667e-05\\
-9.65841936001011	3.33333333333333e-05\\
-9.55776170821467	3e-05\\
-9.45710405641922	2.83333333333333e-05\\
-9.35644640462378	4.5e-05\\
-9.25578875282833	3.66666666666667e-05\\
-9.15513110103289	5e-05\\
-9.05447344923744	5e-05\\
-8.953815797442	4.66666666666667e-05\\
-8.85315814564655	6.33333333333333e-05\\
-8.75250049385111	8.83333333333333e-05\\
-8.65184284205566	8.16666666666667e-05\\
-8.55118519026022	0.000103333333333333\\
-8.45052753846477	9.5e-05\\
-8.34986988666932	0.000128333333333333\\
-8.24921223487388	0.000125\\
-8.14855458307843	0.00012\\
-8.04789693128299	0.000126666666666667\\
-7.94723927948754	0.000171666666666667\\
-7.8465816276921	0.00018\\
-7.74592397589665	0.000213333333333333\\
-7.64526632410121	0.000253333333333333\\
-7.54460867230576	0.000268333333333333\\
-7.44395102051032	0.000248333333333333\\
-7.34329336871487	0.000301666666666667\\
-7.24263571691943	0.00033\\
-7.14197806512398	0.000381666666666667\\
-7.04132041332853	0.000395\\
-6.94066276153309	0.000396666666666667\\
-6.84000510973764	0.000496666666666667\\
-6.7393474579422	0.000515\\
-6.63868980614675	0.000591666666666667\\
-6.53803215435131	0.00068\\
-6.43737450255586	0.000651666666666667\\
-6.33671685076042	0.000845\\
-6.23605919896497	0.000933333333333333\\
-6.13540154716953	0.000951666666666667\\
-6.03474389537408	0.000985\\
-5.93408624357864	0.001095\\
-5.83342859178319	0.00134\\
-5.73277093998774	0.00140666666666667\\
-5.6321132881923	0.00157833333333333\\
-5.53145563639685	0.00177333333333333\\
-5.43079798460141	0.00191333333333333\\
-5.33014033280596	0.002065\\
-5.22948268101052	0.00242333333333333\\
-5.12882502921507	0.00249666666666667\\
-5.02816737741963	0.00283666666666667\\
-4.92750972562418	0.0031\\
-4.82685207382874	0.00341166666666667\\
-4.72619442203329	0.00363666666666667\\
-4.62553677023785	0.004105\\
-4.5248791184424	0.00441\\
-4.42422146664696	0.00499\\
-4.32356381485151	0.00537333333333333\\
-4.22290616305606	0.00580166666666667\\
-4.12224851126062	0.006595\\
-4.02159085946517	0.00678166666666667\\
-3.92093320766973	0.007455\\
-3.82027555587428	0.00838166666666667\\
-3.71961790407884	0.008845\\
-3.6189602522834	0.00978\\
-3.51830260048795	0.0106933333333333\\
-3.4176449486925	0.0113583333333333\\
-3.31698729689706	0.0122783333333333\\
-3.21632964510161	0.0128566666666667\\
-3.11567199330617	0.0138383333333333\\
-3.01501434151072	0.0149433333333333\\
-2.91435668971528	0.0157683333333333\\
-2.81369903791983	0.0163333333333333\\
-2.71304138612438	0.01737\\
-2.61238373432894	0.0180766666666667\\
-2.5117260825335	0.01883\\
-2.41106843073805	0.0193083333333333\\
-2.3104107789426	0.019725\\
-2.20975312714716	0.019985\\
-2.10909547535171	0.0199416666666667\\
-2.00843782355627	0.0194666666666667\\
-1.90778017176082	0.0186916666666667\\
-1.80712251996538	0.01788\\
-1.70646486816993	0.017065\\
-1.60580721637449	0.0152116666666667\\
-1.50514956457904	0.01361\\
-1.4044919127836	0.01178\\
-1.30383426098815	0.00951666666666667\\
-1.2031766091927	0.00753833333333333\\
-1.10251895739726	0.00573\\
-1.00186130560182	0.00401666666666667\\
-0.90120365380637	0.002655\\
-0.800546002010924	0.001455\\
-0.699888350215479	0.000718333333333333\\
-0.599230698420033	0.00031\\
-0.498573046624589	0.00012\\
-0.397915394829145	3e-05\\
-0.297257743033697	3.33333333333333e-06\\
-0.196600091238253	0\\
-0.0959424394428083	0\\
0.0047152123526395	0\\
0.105372864148084	0\\
0.206030515943528	0\\
0.306688167738976	5.16666666666667e-05\\
0.40734581953442	0.002735\\
0.508003471329864	0.0290833333333333\\
0.608661123125309	0.09827\\
0.709318774920757	0.14939\\
0.809976426716201	0.123266666666667\\
0.910634078511645	0.0651783333333333\\
1.01129173030709	0.02357\\
1.11194938210254	0.00675\\
1.21260703389798	0.0014\\
1.31326468569343	0.000255\\
1.41392233748887	4e-05\\
1.51457998928432	8.33333333333333e-06\\
}; \addlegendentry{$|\X|=30, |\Sen|=20$};
\draw[decorate,decoration={brace,mirror,amplitude=7}]
([yshift=-20pt]axis cs:-15.5,0) --
node[below=5pt] {\huge$\nu(x)$} 
([yshift=-20pt]axis cs:0,0);
\draw[decorate,decoration={brace,mirror,amplitude=7}]
([yshift=-20pt]axis cs:0,0) --
node[below=5pt] {\huge  $\xi(x)$} 
([yshift=-20pt]axis cs:1.7,0);

\end{axis}
\end{tikzpicture}%
}
		\caption{Histogram of $\nu(x)=\min_{s}\lglif$ and $\xi(x)=\max_{s}\lglif$ for $10^4$ randomly generated distributions, with $|\X|=30$, $|\Sen|=20$.}
		\label{fig:density}
	\end{figure}

	The existing literature has tried to minimize both $|\nu(x)|$ and $\xi(x)$ or restrict them below a given value. 
	The privacy algorithms in previous works apply the same bound for  $|\nu(x)|$ and $\xi(x)$. 
	We call this scenario symmetric local information privacy (SLIP).
	However, typically, the range of values for $\nu(x)$ and $\xi(x)$ are very different.
	Fig.~\ref{fig:density} shows the histogram of $\nu(x)$ and $\xi(x)$.
	We observe that the range of $\nu(x)$  [-15.5,-0.5] is much larger than the range of $\xi(x)$, [0.3,1.5].
	Moreover, Fig.~\ref{fig:density} shows larger probability values for $\xi(x)$ than $\nu(x)$. 
	Similarly, stemming from the non-negativity of mutual information $I(S;X)$, $P_{S,X}$ is typically much smaller for a very negative $\nu(x)$ than for a very positive $\xi(x)$.
	Therefore, it seems an unnecessary restriction to apply the same bounds to both of these quantities.

	In this work, we generalize LIP by designation of different privacy budgets, $\epsl$ and $\epsu$, for restricting $\nu(x)$ and $\xi(x)$, respectively; and we call it asymmetric local information privacy (ALIP).
	Then, we investigate the privacy-utility trade-off for ALIP by applying it to the watchdog privacy mechanism.
	We demonstrate that in ALIP, only max-lift affects the bounds on IT average measures including, MI, maximal leakage, $\alpha$-leakage (Arimoto MI), and Sibson MI; and LDP is upper bounded  by $\eps=\epsu+\epsl$. 
   We show that for a fixed $\eps$, by relaxation of $\nu(x)$ in the watchdog mechanism, ALIP ($\epsl>\epsu$) can not only enhance the utility compared to SLIP ($\epsl=\epsu$), but also provide a tighter upper bound $\epsu$ on the aforementioned IT measures.  
%%%%%%%%%%%%%%%%%%%%%%%%%%%%%%%%%%%%%%%%%%%%%%%%%%%%%%%%%%%%%%%%%%%%%%%%%%%%%%%%%%%%%%%%%%%%%%%%%%%%%%%%%%%%%%%%%%%%%%%%%%%%%%%%%%%%%%%%%%%%%%%%%%%%%%%%%%%%%%%%%%%%%%%%%%%%%%%%%%		
\section{Asymmetric local information privacy and the Watchdog Mechanism}

	Denote useful data by $X$ with alphabet $\X$ and confidential features correlated with $X$ by  $S$ with alphabet $\Sen$; they are correlated via joint distribution $(S,X) \sim \PSX$. 
	Our goal is to publish a sanitized version of $X$, denoted by $Y$ with alphabet $\Y$, to protect the privacy of $S$ and provide appropriate statistical utility for $X$. 
	They form a Markov chain $S-X-Y$ where $P_{Y|S,X}(y|s,x)=\PYgX(y|x)$ for all $s,x,y$, and $\PYgX(y|x)$ is the privacy mechanism.
	\subsection{Asymmetric Local Information Privacy}
	\begin{definition}\label{Def:ALIP} 
		For a given useful data $X$ and private data $S$ where $(S,X)\sim \PSX$, the privacy mechanism $\PYgX$ satisfies $(\epsl,\epsu)$-ALIP for some $\epsl, \epsu \in \RealP$ if  $\forall s,y$ :
		\begin{equation}\label{eq:ALIP}
			-\epsl \leq i(s,y) \leq \epsu.
		\end{equation}
		 When \eqref{eq:ALIP} holds we say $Y$ is an $(\epsl,\epsu)$-ALIP private version of $S$.
	\end{definition}
	For instance, an adversary can eliminate symbols $x$ with very small lift values from the list of likely sensitive features or prioritize symbols $x$ that enhance the likelihood of a certain $s$ happening. 
	% %
	% Thus, for each $x$, $\min_{s}\lglif$ and $\max_{s}\lglif$ determine the extreme risk of publishing $x$ and should be restricted to protect the privacy of sensitive features. 
	\begin{proposition}\label{prop:ALIP properties}
		When $Y$ is $(\epsl,\epsu)$-ALIP private version of $S$, the following properties are held \cite{2019Watchdog}:
		\begin{enumerate}
			\item \label{prop:LDP} $\PYgS$ is $(\epsl+\epsu)$-locally differential private, i.e.
			\begin{equation}
				\sup_{\forall y, s, s'}\frac{\PYgS(y|s)}{\PYgS(y|s')} \leq e^{(\epsl+\epsu)} .
			\end{equation}
			\item \label{prop:MI} The $\alpha$-lift, mutual information $I(S;Y$), and maximal leakage between $S$ and $Y$, are upper bounded by $\epsu$.
			\item \label{prop:Sibson} The Sibson MI $I_{\alpha}^{S}(S;Y)$ and Arimoto MI $I_{\alpha}^{A}(S;Y)$ are upper bounded by $\frac{\alpha}{\alpha-1}\epsu$.
		\end{enumerate}
	\end{proposition}
	\noindent See Appendices \ref{app:definitions} and \ref{app:proof of prop 1}  for the proof, which follows the proof of \cite[Proposition~1]{2019Watchdog}.
	
	\vspace{3pt}
	Proposition \ref{prop:LDP}-\ref{prop:ALIP properties}) means that if $\eps=\epsl+\epsu$ then $\eps$-LDP is achieved. In other words, LIP methods could be applied to attain LDP.
	Here, for a fixed $\eps$, different values of $\epsu$ and $\epsl$ can be considered to provide different ALIP scenarios, and SLIP ($\epsl=\epsu$) is just one special case.
	We will show in Section \ref{sec:numerical results} that for a fixed $\eps$, ALIP enhances utility  compared to SLIP when $\epsl>\epsu$.
	This is an interesting property where we can enhance utility for a fixed LDP constraint and at the same time, we can decrease the guessing ability of an adversary by enforcing a more strict bound on $\xi(x)$.
	Propositions \ref{prop:MI}-\ref{prop:ALIP properties}) and \ref{prop:Sibson}-\ref{prop:ALIP properties}) show that $\alpha$-lift, mutual information and other mentioned IT measures are upper bounded by $\epsu$. When $\epsl>\epsu$, this means that ALIP can not only improve utility, but also provide a tighter upper bound on other IT measures than SLIP.
	
	It should be noted that the watchdog-based utility enhancements in \cite{2020SadeghiITW} and \cite{2021enhancingzarrabian} applied equal bounds on max-lift and min-lift. \cite{2020SadeghiITW} relaxed privacy constraints, i.e., allowing some probability of breaching the bounds on lifts; \cite{2021enhancingzarrabian} required the search of a locally optimal subset partition of $X_H$. However, we generalize the definition of LIP to match the framework with asymmetric property of the log-lift.
	 
\subsection{Application of ALIP to the Watchdog Mechanism}
	The watchdog privacy mechanism has been proposed to achieve LIP in \cite{2019Watchdog}. Here, we apply ALIP to it and define the asymmetric privacy watchdog.
	
	\begin{definition}
		\textit{{Asymmetric watchdog privacy mechanism}}: For a given $(X,S) \sim \PSX$ and $\epsl,\epsu \in \RealP$, the watchdog mechanism bi-partitions $\X$ into subsets of low-risk and high-risk symbols $\X_L$ and $\X_H$, respectively, as follows:
		\begin{equation}
			\begin{aligned}
				\X_L\triangleq \{x\in \X: & \quad  -\epsl \leq i(s,x) \leq \epsu, \quad \forall s\in \Sen \},\\
				&\X_H=\X \setminus \X_L;
			\end{aligned} 
		\end{equation}
		where $x \in \X_L$ are published without perturbation and $x \in \X_H$ are randomized via a randomization $r_{Y|X}(y|x)$. As a result, the privacy mechanism $\PYgX$ is given by:
	\begin{equation} \label{eq:origina watchdog}
		\PYgX(y|x) = \begin{cases} 
			\mathbf{1}_{\{x=y\}} & x,y \in \X_L,\\
			r_{Y|X}(y|x)		 & x,y \in \X_H,\\
			0					 & \textup{otherwise};
		\end{cases}
	\end{equation}
	where \small$\mathbf{1}_{\{x=y\}}$ \normalsize  is the indicator function and \small $\displaystyle\sum_{y\in\X_H}r_{Y|X}(y|x)=1$.\normalsize
	\end{definition}
 
	Similar to SLIP \cite{2020SadeghiITW}, the optimal $r_{Y|X}(y|x)$ which minimizes the privacy leakage in $\X_H$ is an $X$-invariant randomization $R_{Y}(y)$ that only depends on $y\in \X_H$, which is constant for all $x \in \X_H$, and is zero otherwise. 
	An instance of $R_{Y}(y)$ is uniform randomization, $R_{Y}(y)=\frac{1}{|\X_H|},\hspace{2pt} y \in \X_H$. The other option is \textit{complete merging} which is applied in this paper where all $x \in \X_H$ are mapped to only one super symbol $y^* \in \X_H$ where $R_{Y}(y^*)=1$, and  $R_{Y}(y)=0, y \neq y^*.$ 

	\begin{proposition}\label{prop:Xinvarinat}
		For a given $(\epsu,\epsl)$ and $\{\X_L,\X_H\}$, X-invariant randomization $R_{Y}(y)$ minimizes  the privacy leakage in $\X_H$. 
		The minimum achievable log-lift upper bound $\epsu^{*}$ and lower bound $\epsl^{*}$ in $\X_H$, respectively, are given by:
	\begin{align}
			&\epsu^{*}=\max_{s}i(s,\X_H):=\max_{s}\log\frac{P(\X_H|s)}{P(\X_H)},\\
			&\epsl^{*}=\left|\min_{s}i(s,\X_H)\right|:=\left|\min_{s}\log\frac{P(\X_H|s)}{P(\X_H)}\right|,
	\end{align}
	where \small $\displaystyle P(\X_H|s)=\sum_{x \in \X_H}\PXgS(x|s)$ \normalsize, and \small$\displaystyle P(\X_H)=\sum_{x \in \mathcal{X}_H}\PX(x)$.\normalsize
	\end{proposition}
	\noindent {The proof is similar to the proof of \cite[Corollary 2]{2020SadeghiITW}, however, for the sake of completeness it is given in Appendix \ref{app:proof of prop 2}.}

	\subsection{Utility Measure}
	To measure utility, we use the normalized mutual information (NMI). Mutual information between $X$ and $Y$ in the watchdog mechanism is given by \cite{2020SadeghiITW}:
	\begin{equation}
		I(X;Y)=H(X)+\sum_{x \in \mathcal{X}_H} \PX(x) \log \frac{\PX(x)}{\PX\left(\mathcal{X}_H\right)}.
	\end{equation}
 	Then, the NMI will be 
	\begin{equation}
		\text{NMI}=\frac{I(X;Y)}{H(X)} \in [0,1].
	\end{equation}

%%%%%%%%%%%%%%%%%%%%%%%%%%%%%%%%%%%%%%%%%%%%%%%%%%%%%%%%%%%%%%%%%%%%%%%%%%%%%%%%%%%%%%%%%%%%%%%%%%%%%%%%%%%%%%%%%%%%%%%%%%%%%%%%%%%%%%%%%%%%%%%%%%%%%%%%%%%%%%%%%%%%%%%%%%%%%%%%%%
\section{Numerical results}\label{sec:numerical results}
	In this section, we investigate the properties of ALIP numerically. All results have been derived from $10^4$ randomly generated distributions $\PSX$ under the asymmetric watchdog mechanism where $|\X|=30$, $|\Sen|=20$. 
\subsection{Privacy-Utility Trade-off}
	In this section, we demonstrate utility for different values of $\epsu$ and $\epsl$.
	In Fig.~\ref{fig:UtilityStaircase}, the NMI, is shown for different values of $\epsu$ and $\epsl$, under the asymmetric watchdog mechanism. 
	Here, we derived the average utility of $10^4$ randomly generated $\PSX$ where ${\epsu}=\{0.6, 0.8, 1\}$ and ${\epsl}=\{0.4,0.5,0.6,\cdots,6\}$.
	For each $\epsu$, we increase $\epsl$ to demonstrate the privacy-utility trade-off. 
	For $\epsu=0.6$, the highest value of utility is $0.35$, which is a low value, due to a very strict condition on $\xi(x)$. 
	As it is observed, there is a greater gap between $\epsu=0.6$ and $\epsu=0.8$ and a smaller gap to $\epsu=1$. 
	The reason is that the values of $\xi(x)$ belong to $[0.6,1]$ in our experiments and most of the values range between $0.6$ and $0.8$.
	 Another observation is that the NMI gets mostly saturated after a certain point when increasing $\epsl$.
	 It is due to the fact that increasing $\epsl$ does not increase utility by much as it is already limited by $\epsu$. 
	\begin{figure}[]
		\centering
		\scalebox{0.5}{\definecolor{mycolor1}{rgb}{0,0.447,0.741}%
\definecolor{mycolor2}{rgb}{0.85,0.325,0.098}%
\definecolor{mycolor3}{rgb}{0.494,0.184,0.556}%
\definecolor{mycolor4}{rgb}{0.929,0.694,0.125}%

\begin{tikzpicture}
\begin{axis}[
width=4.854in,
height=3in,
scale only axis,
separate axis lines,
every outer x axis line/.append style={darkgray!60!black},
every x tick label/.append style={font=\color{darkgray!60!black},font=\Large},
xmin=0,
xmax=6,
xlabel={\Huge $\epsilon_{l}$},
xmajorgrids,
every outer y axis line/.append style={darkgray!60!black},
every y tick label/.append style={font=\color{darkgray!60!black},font=\Large},
ymin=0,
ymax=1,
ylabel={\Large NMI},
ymajorgrids,
legend style={at={(0.24,0.98)},draw=darkgray!60!black,fill=white,legend cell align=left, font=\Large},
]
\addplot [
color=mycolor1,
solid,
line width=2.0pt,
mark=pentagon,
mark options={solid},
mark size=2pt,
]
table[row sep=crcr]{
0.5 6.34343425749016e-05\\
0.6 0.000308139643120076\\
0.7 0.00103014450195466\\
0.8 0.00264443721016706\\
0.9 0.00553386899745179\\
1 0.0099359138654853\\
1.1 0.0163039454442379\\
1.2 0.0247736034410441\\
1.3 0.0344992614991522\\
1.4 0.0471281428445606\\
1.5 0.0605004429593265\\
1.6 0.0745582805286553\\
1.7 0.0894530900602646\\
1.8 0.105184218287227\\
1.9 0.120381702382709\\
2 0.13527328906182\\
2.1 0.150308294224719\\
2.2 0.164350854745629\\
2.3 0.178007471097596\\
2.4 0.190394748262642\\
2.5 0.203291942981999\\
2.6 0.214636405184952\\
2.7 0.225210242411062\\
2.8 0.235128572103727\\
2.9 0.24495131438835\\
3 0.253581908813891\\
3.1 0.261944598608787\\
3.2 0.269501521331705\\
3.3 0.276854214495467\\
3.4 0.282828376646822\\
3.5 0.288933574563857\\
3.6 0.294517605647378\\
3.7 0.29973259566417\\
3.8 0.304291881168103\\
3.9 0.308354033694472\\
4 0.312281739110677\\
4.1 0.315809241029188\\
4.2 0.31913493268579\\
4.3 0.321984502903378\\
4.4 0.324528966620772\\
4.5 0.327071765625806\\
4.6 0.329315334503828\\
4.7 0.331220253463822\\
4.8 0.332953471132914\\
4.9 0.334449757783878\\
5 0.335915725077417\\
5.1 0.337126720193811\\
5.2 0.338249557065895\\
5.3 0.339377637285403\\
5.4 0.340499486197871\\
5.5 0.341332588287188\\
5.6 0.342142413824323\\
5.7 0.342703251187196\\
5.8 0.343483164115352\\
5.9 0.344067826946594\\
6 0.344538352314747\\
}; \addlegendentry{$\epsu=0.6$};
\addplot [
color=mycolor2,
solid,
line width=2.0pt,
mark=x,
mark options={solid},
mark size=3pt,
]
table[row sep=crcr]{
0.5 7.92835371652719e-05\\
0.6 0.000333502093115647\\
0.7 0.00125059756152875\\
0.8 0.0033077224314659\\
0.9 0.00744315315637299\\
1 0.0140737875697031\\
1.1 0.0244588443219437\\
1.2 0.0395093500571313\\
1.3 0.0583349904218068\\
1.4 0.0837045884747269\\
1.5 0.111949834428441\\
1.6 0.144003828022931\\
1.7 0.178931786795249\\
1.8 0.217237775124271\\
1.9 0.255882456012896\\
2 0.295344492019525\\
2.1 0.335583628448725\\
2.2 0.37451914865558\\
2.3 0.412765127503398\\
2.4 0.449726025353914\\
2.5 0.486576165600663\\
2.6 0.520576763224999\\
2.7 0.553009691227151\\
2.8 0.583514991378865\\
2.9 0.61220764358627\\
3 0.638701722624697\\
3.1 0.662867787461011\\
3.2 0.685676319083216\\
3.3 0.706565655190769\\
3.4 0.725115997430514\\
3.5 0.742609060084507\\
3.6 0.758587052195561\\
3.7 0.773529441912562\\
3.8 0.786423638091863\\
3.9 0.798561121604873\\
4 0.809272151657781\\
4.1 0.819056408391419\\
4.2 0.827978668857836\\
4.3 0.836022772094071\\
4.4 0.843172606860692\\
4.5 0.849609984762093\\
4.6 0.855662739110794\\
4.7 0.86096271355255\\
4.8 0.865533089609645\\
4.9 0.869668619293289\\
5 0.873690276689295\\
5.1 0.877098181328177\\
5.2 0.880102422507584\\
5.3 0.883221510739226\\
5.4 0.885883170342124\\
5.5 0.888144189781023\\
5.6 0.89033124029949\\
5.7 0.892091536790115\\
5.8 0.893813881177116\\
5.9 0.895255532954126\\
6 0.89653539043932\\
}; \addlegendentry{$\epsu=0.8$};
\addplot [
color=mycolor3,
solid,
line width=2.0pt,
mark=oplus,
mark options={solid},
mark size=3pt,
]
table[row sep=crcr]{
0.5 7.92835371652719e-05\\
0.6 0.000342581355059778\\
0.7 0.00125967682347288\\
0.8 0.00338468411734281\\
0.9 0.00758794391134086\\
1 0.0144416619633345\\
1.1 0.0252229932634283\\
1.2 0.0410670579291986\\
1.3 0.061110231226525\\
1.4 0.0884245558053774\\
1.5 0.119331508638243\\
1.6 0.154530464935676\\
1.7 0.193624333712263\\
1.8 0.236542261840243\\
1.9 0.280577074925125\\
2 0.32584289733151\\
2.1 0.372024421712674\\
2.2 0.416957802326922\\
2.3 0.462153297332712\\
2.4 0.505298868357341\\
2.5 0.548252919455574\\
2.6 0.588145707801183\\
2.7 0.626058467455412\\
2.8 0.662081484875211\\
2.9 0.695739235942329\\
3 0.726712200517206\\
3.1 0.754856429366282\\
3.2 0.781082324170607\\
3.3 0.805187297949868\\
3.4 0.826460816342824\\
3.5 0.846102489579004\\
3.6 0.864115110160495\\
3.7 0.880539094717808\\
3.8 0.894273148768118\\
3.9 0.907259463213927\\
4 0.918474764731076\\
4.1 0.928647551081412\\
4.2 0.937593923362452\\
4.3 0.945458364522294\\
4.4 0.952259004239955\\
4.5 0.95826763896196\\
4.6 0.96372706233269\\
4.7 0.968394980384121\\
4.8 0.972170853395646\\
4.9 0.975580087425402\\
5 0.978762968657842\\
5.1 0.981358174107466\\
5.2 0.983541593425965\\
5.3 0.985723973015794\\
5.4 0.9874409938649\\
5.5 0.988903846566153\\
5.6 0.990266889178254\\
5.7 0.991266609228142\\
5.8 0.992283159738341\\
5.9 0.993051700777612\\
6 0.993726377381761\\
}; \addlegendentry{$\epsu=1$};
%\addplot [
%color=mycolor4,
%solid,
%line width=2.0pt,
%mark=diamond,
%mark options={solid},
%mark size=2pt,
%]
%table[row sep=crcr]{
%0.5 7.92835371652719e-05\\
%0.6 0.000342581355059778\\
%0.7 0.00125967682347288\\
%0.8 0.00338468411734281\\
%0.9 0.00758794391134086\\
%1 0.0144481859805555\\
%1.1 0.0252370320429803\\
%1.2 0.0411114097961488\\
%1.3 0.0612049411462415\\
%1.4 0.0886140097507893\\
%1.5 0.119695113816224\\
%1.6 0.155018959972295\\
%1.7 0.194370712774312\\
%1.8 0.237557371185655\\
%1.9 0.28192461227599\\
%2 0.327616636220167\\
%2.1 0.374109231090922\\
%2.2 0.41941194042599\\
%2.3 0.465161505936169\\
%2.4 0.508774123701438\\
%2.5 0.552269089628555\\
%2.6 0.592528646537217\\
%2.7 0.630986399422478\\
%2.8 0.667556374260705\\
%2.9 0.701706137993415\\
%3 0.733097425808101\\
%3.1 0.761541708195655\\
%3.2 0.788046640784151\\
%3.3 0.812405000936383\\
%3.4 0.833913628448283\\
%3.5 0.853646656760691\\
%3.6 0.871763534509774\\
%3.7 0.888262419595587\\
%3.8 0.902007604620757\\
%3.9 0.914986715814785\\
%4 0.926121410068351\\
%4.1 0.936219567799234\\
%4.2 0.945069532652355\\
%4.3 0.952869250088704\\
%4.4 0.959480946379108\\
%4.5 0.965336096501668\\
%4.6 0.970559088494896\\
%4.7 0.97503700090005\\
%4.8 0.978605428932316\\
%4.9 0.981834817061768\\
%5 0.984759768344822\\
%5.1 0.987123580602616\\
%5.2 0.989037373524925\\
%5.3 0.99095539784502\\
%5.4 0.992464375333822\\
%5.5 0.993669078173017\\
%5.6 0.994786553256997\\
%5.7 0.995566917924104\\
%5.8 0.996302495991755\\
%5.9 0.996900425087935\\
%6 0.997406818592156\\
%}; \addlegendentry{$\epsu=1.2$};
\end{axis}
\end{tikzpicture}%
}
		\caption{Average utility of $10^4$ randomly generated distributions $\PSX$ measured by NMI under the watchdog mechanism for different values of $\epsu$ and $\epsl$.}
		\label{fig:UtilityStaircase}
	\end{figure}
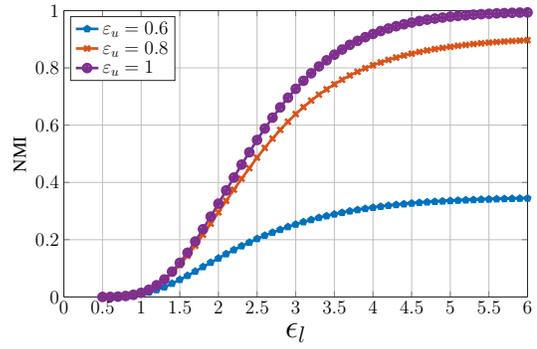
	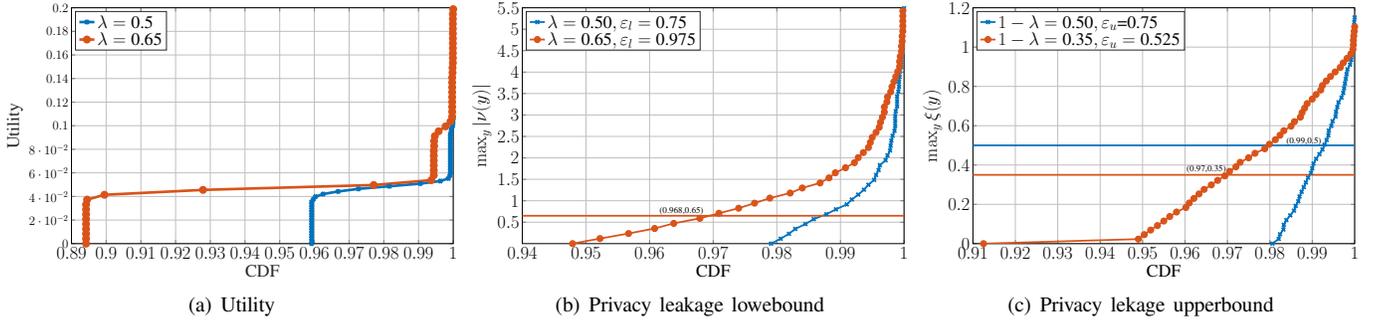
\begin{figure*}
		\centering     %%% not \center
		\subfigure[\label{fig:eps1utility} Utility]{\scalebox{0.333}{\definecolor{mycolor1}{rgb}{0,0.447,0.741}%
\definecolor{mycolor2}{rgb}{0.85,0.325,0.098}%
\begin{tikzpicture}

\begin{axis}[%
width=6in,
height=3.71in,
at={(0.758in,0.516in)},
scale only axis,
xmin=0.89,
xmax=1,
xlabel style={font=\color{white!15!black}},
xlabel={\huge CDF},
every x tick label/.append style={font=\color{darkgray!60!black},font=\huge},
ymin=0,
ymax=0.2,
ylabel style={font=\color{white!15!black}},
ylabel={\huge Utility},
every y tick label/.append style={font=\color{darkgray!60!black},font=\Large},
axis background/.style={fill=white},
title style={font=\bfseries},
xmajorgrids,
ymajorgrids,
legend style={at={(0.01,0.82)}, anchor=south west, legend cell align=left, align=left, draw=white!15!black,font=\huge}
]
\addplot [color=mycolor1, line width=3.0pt, mark=x,mark options={solid},mark size=3pt]
  table[row sep=crcr]{%
0.9592	-6.66133814775094e-16\\
0.9592	0.00221134197029687\\
0.9592	0.00442268394059441\\
0.9592	0.00663402591089195\\
0.9592	0.00884536788118948\\
0.9592	0.011056709851487\\
0.9592	0.0132680518217846\\
0.9592	0.0154793937920821\\
0.9592	0.0176907357623796\\
0.9592	0.0199020777326772\\
0.9592	0.0221134197029747\\
0.9592	0.0243247616732722\\
0.9592	0.0265361036435698\\
0.9592	0.0287474456138673\\
0.9592	0.0309587875841649\\
0.9592	0.0331701295544624\\
0.9592	0.0353814715247599\\
0.9597	0.0375928134950575\\
0.9603	0.039804155465355\\
0.9625	0.0420154974356525\\
0.9668	0.0442268394059501\\
0.9727	0.0464381813762476\\
0.9816	0.0486495233465452\\
0.9905	0.0508608653168427\\
0.9963	0.0530722072871402\\
0.9986	0.0552835492574378\\
0.999	0.0574948912277353\\
0.9992	0.0597062331980328\\
0.9992	0.0619175751683304\\
0.9992	0.0641289171386279\\
0.9992	0.0663402591089255\\
0.9992	0.068551601079223\\
0.9992	0.0707629430495205\\
0.9992	0.0729742850198181\\
0.9992	0.0751856269901156\\
0.9992	0.0773969689604131\\
0.9992	0.0796083109307107\\
0.9992	0.0818196529010082\\
0.9992	0.0840309948713058\\
0.9992	0.0862423368416033\\
0.9992	0.0884536788119008\\
0.9992	0.0906650207821984\\
0.9992	0.0928763627524959\\
0.9994	0.0950877047227934\\
0.9995	0.097299046693091\\
0.9998	0.0995103886633885\\
0.9999	0.101721730633686\\
0.9999	0.103933072603984\\
1	0.106144414574281\\
};
\addlegendentry{$\lambda=\text{0.5}$}

\addplot [color=mycolor2, line width=3.0pt,mark=oplus,mark options={solid},mark size=3pt]
  table[row sep=crcr]{%
0.8942	-6.66133814775094e-16\\
0.8942	0.00414194206759208\\
0.8942	0.00828388413518483\\
0.8942	0.0124258262027776\\
0.8942	0.0165677682703703\\
0.8942	0.0207097103379631\\
0.8942	0.0248516524055558\\
0.8942	0.0289935944731486\\
0.8942	0.0331355365407413\\
0.8945	0.0372774786083341\\
0.8995	0.0414194206759268\\
0.9279	0.0455613627435195\\
0.9771	0.0497033048111123\\
0.9937	0.053845246878705\\
0.9944	0.0579871889462978\\
0.9944	0.0621291310138905\\
0.9944	0.0662710730814833\\
0.9944	0.070413015149076\\
0.9944	0.0745549572166688\\
0.9944	0.0786968992842615\\
0.9945	0.0828388413518543\\
0.9945	0.086980783419447\\
0.9947	0.0911227254870398\\
0.9958	0.0952646675546325\\
0.9978	0.0994066096222253\\
0.9992	0.103548551689818\\
0.9997	0.107690493757411\\
0.9997	0.111832435825004\\
0.9997	0.115974377892596\\
0.9997	0.120116319960189\\
0.9997	0.124258262027782\\
0.9997	0.128400204095374\\
0.9997	0.132542146162967\\
0.9997	0.13668408823056\\
0.9997	0.140826030298153\\
0.9998	0.144967972365745\\
0.9998	0.149109914433338\\
0.9999	0.153251856500931\\
0.9999	0.157393798568524\\
0.9999	0.161535740636116\\
0.9999	0.165677682703709\\
0.9999	0.169819624771302\\
0.9999	0.173961566838895\\
0.9999	0.178103508906487\\
0.9999	0.18224545097408\\
0.9999	0.186387393041673\\
0.9999	0.190529335109266\\
0.9999	0.194671277176858\\
1	0.198813219244451\\
};
\addlegendentry{$\lambda=\text{0.65}$}

\end{axis}

\begin{axis}[%
width=5.833in,
height=4.375in,
at={(0in,0in)},
scale only axis,
xmin=0,
xmax=1,
ymin=0,
ymax=1,
axis line style={draw=none},
ticks=none,
axis x line*=bottom,
axis y line*=left
]
\end{axis}
\end{tikzpicture}%
}}
		\subfigure[\label{fig:eps1leakagel} Privacy leakage lowebound]{\scalebox{0.333}{\definecolor{mycolor1}{rgb}{0,0.447,0.741}%
\definecolor{mycolor2}{rgb}{0.85,0.325,0.098}%
\begin{tikzpicture}

\begin{axis}[%
width=6in,
height=3.71in,
at={(0.758in,0.481in)},
scale only axis,
xmin=0.94,
xmax=1,
xtick={0.94,0.95,0.96,0.97,0.98,0.99,1},
every x tick label/.append style={font=\color{darkgray!60!black},font=\huge},
xlabel={\huge CDF},
ymin=0,
ymax=5.5,
ylabel={\huge $\max_{y}|\nu(y)|$},
ytick={0.5,1, 1.5,2,2.5,3,3.5,4,4.5,5,5.5},
every y tick label/.append style={font=\color{darkgray!60!black},font=\huge},
axis background/.style={fill=white},
xmajorgrids,
ymajorgrids,
legend style={at={(0.01,0.82)}, anchor=south west, legend cell align=left, align=left, draw=white!15!black,font=\huge}
]
\addplot [color=mycolor1, line width=2.0pt, mark=x,mark options={solid},mark size=3pt]
  table[row sep=crcr]{%
0.9791	0\\
0.9806	0.114360603824742\\
0.9819	0.228721207649485\\
0.9829	0.343081811474227\\
0.9845	0.457442415298969\\
0.9858	0.571803019123711\\
0.9878	0.686163622948454\\
0.9893	0.800524226773196\\
0.991	0.914884830597938\\
0.9918	1.02924543442268\\
0.9928	1.14360603824742\\
0.9937	1.25796664207217\\
0.9944	1.37232724589691\\
0.9953	1.48668784972165\\
0.9956	1.60104845354639\\
0.996	1.71540905737113\\
0.9962	1.82976966119588\\
0.9971	1.94413026502062\\
0.9974	2.05849086884536\\
0.9978	2.1728514726701\\
0.9979	2.28721207649485\\
0.998	2.40157268031959\\
0.9982	2.51593328414433\\
0.9985	2.63029388796907\\
0.9986	2.74465449179381\\
0.9986	2.85901509561856\\
0.9986	2.9733756994433\\
0.9987	3.08773630326804\\
0.9989	3.20209690709278\\
0.9989	3.31645751091753\\
0.9989	3.43081811474227\\
0.9991	3.54517871856701\\
0.9992	3.65953932239175\\
0.9992	3.7738999262165\\
0.9993	3.88826053004124\\
0.9993	4.00262113386598\\
0.9996	4.11698173769072\\
0.9996	4.23134234151546\\
0.9996	4.34570294534021\\
0.9998	4.46006354916495\\
0.9998	4.57442415298969\\
0.9998	4.68878475681443\\
0.9998	4.80314536063918\\
0.9998	4.91750596446392\\
0.9999	5.03186656828866\\
0.9999	5.1462271721134\\
0.9999	5.26058777593814\\
0.9999	5.37494837976289\\
1	5.48930898358763\\
};\addlegendentry{$\lambda=\text{0.50},  \epsl=\text{0.75}$}
\addplot [color=mycolor2, line width=2.0pt,mark=oplus,mark options={solid},mark size=3pt]
  table[row sep=crcr]{%
0.9479	0\\
0.9522	0.117976412758102\\
0.9567	0.235952825516204\\
0.9608	0.353929238274306\\
0.9638	0.471905651032408\\
0.9679	0.58988206379051\\
0.9709	0.707858476548612\\
0.974	0.825834889306713\\
0.9765	0.943811302064815\\
0.9789	1.06178771482292\\
0.9821	1.17976412758102\\
0.984	1.29774054033912\\
0.9868	1.41571695309722\\
0.9882	1.53369336585533\\
0.9892	1.65166977861343\\
0.9908	1.76964619137153\\
0.9922	1.88762260412963\\
0.9928	2.00559901688773\\
0.9936	2.12357542964583\\
0.9944	2.24155184240394\\
0.9947	2.35952825516204\\
0.995	2.47750466792014\\
0.9956	2.59548108067824\\
0.9961	2.71345749343634\\
0.9963	2.83143390619445\\
0.9966	2.94941031895255\\
0.9969	3.06738673171065\\
0.9969	3.18536314446875\\
0.9973	3.30333955722685\\
0.9974	3.42131596998496\\
0.9979	3.53929238274306\\
0.9981	3.65726879550116\\
0.9984	3.77524520825926\\
0.998699999999999	3.89322162101736\\
0.998999999999999	4.01119803377547\\
0.999299999999999	4.12917444653357\\
0.999299999999999	4.24715085929167\\
0.999399999999999	4.36512727204977\\
0.999599999999999	4.48310368480787\\
0.999599999999999	4.60108009756598\\
0.999699999999999	4.71905651032408\\
0.999699999999999	4.83703292308218\\
0.999799999999999	4.95500933584028\\
0.999799999999999	5.07298574859838\\
0.999799999999999	5.19096216135648\\
0.999799999999999	5.30893857411459\\
0.999799999999999	5.42691498687269\\
0.999899999999999	5.54489139963079\\
0.999999999999999	5.66286781238889\\
};\addlegendentry{$\lambda=\text{0.65}, \epsl=\text{0.975}$}

%-------------------------------------------------------------------------------------------
\addplot [mark=none, color=mycolor2, line width=2.0pt] coordinates {(0.94,0.65) (1,0.65)};
\node (a) at (0.965,0.76) {(0.968,0.65)};
%-------------------------------------------------------------------------------------------
\end{axis}

\begin{axis}[%
width=5.833in,
height=4.375in,
at={(0in,0in)},
scale only axis,
xmin=0,
xmax=1,
ymin=0,
ymax=1,
axis line style={draw=none},
ticks=none,
axis x line*=bottom,
axis y line*=left
]
\end{axis}
\end{tikzpicture}%
}}
		\subfigure[\label{fig:eps1leakageu} Privacy lekage upperbound]{\scalebox{0.333}{\definecolor{mycolor1}{rgb}{0,0.447,0.741}%
\definecolor{mycolor2}{rgb}{0.85,0.325,0.098}%
\begin{tikzpicture}

\begin{axis}[%
width=6in,
height=3.71in,
at={(0.758in,0.481in)},
scale only axis,
xmin=0.91,
xmax=1,
xlabel={\huge CDF},
every x tick label/.append style={font=\color{darkgray!60!black},font=\huge},
ymin=0,
ymax=1.2,
every y tick label/.append style={font=\color{darkgray!60!black},font=\huge},
ylabel={\huge $\max_{y}\xi(y)$},
axis background/.style={fill=white},
xmajorgrids,
ymajorgrids,
legend style={at={(0.01,0.82)}, anchor=south west, legend cell align=left, align=left, draw=white!15!black,font=\huge}
]
\addplot [color=mycolor1, line width=2.0pt,mark=x,mark options={solid},mark size=3pt]
  table[row sep=crcr]{%
0.9806	-4.44089209850063e-16\\
0.9821	0.0239668880012835\\
0.9824	0.0479337760025674\\
0.9832	0.0719006640038513\\
0.9833	0.0958675520051353\\
0.9839	0.119834440006419\\
0.9848	0.143801328007703\\
0.9857	0.167768216008987\\
0.9861	0.191735104010271\\
0.9867	0.215701992011555\\
0.987	0.239668880012839\\
0.9875	0.263635768014123\\
0.988	0.287602656015407\\
0.9885	0.311569544016691\\
0.989	0.335536432017975\\
0.9897	0.359503320019258\\
0.99	0.383470208020542\\
0.9903	0.407437096021826\\
0.9909	0.43140398402311\\
0.9917	0.455370872024394\\
0.9924	0.479337760025678\\
0.9929	0.503304648026962\\
0.9934	0.527271536028246\\
0.9938	0.55123842402953\\
0.9939	0.575205312030814\\
0.9945	0.599172200032098\\
0.9948	0.623139088033382\\
0.9955	0.647105976034666\\
0.9956	0.67107286403595\\
0.9959	0.695039752037233\\
0.9962	0.719006640038517\\
0.9969	0.742973528039801\\
0.9973	0.766940416041085\\
0.9975	0.790907304042369\\
0.9977	0.814874192043653\\
0.9979	0.838841080044937\\
0.9981	0.862807968046221\\
0.9983	0.886774856047505\\
0.9989	0.910741744048789\\
0.9992	0.934708632050073\\
0.9993	0.958675520051357\\
0.999599999999999	0.982642408052641\\
0.999599999999999	1.00660929605392\\
0.999599999999999	1.03057618405521\\
0.999599999999999	1.05454307205649\\
0.999799999999999	1.07850996005778\\
0.999899999999999	1.10247684805906\\
0.999899999999999	1.12644373606034\\
0.999999999999999	1.15041062406163\\
};\addlegendentry{$1-\lambda=\text{0.50},  \epsu \text{=0.75}$}

\addplot [color=mycolor2, line width=2.0pt,mark=oplus,mark options={solid},mark size=3pt]
  table[row sep=crcr]{%
0.9125	-4.44089209850063e-16\\
0.949	0.0229862499221709\\
0.9504	0.0459724998443423\\
0.9519	0.0689587497665137\\
0.9535	0.0919449996886851\\
0.955	0.114931249610857\\
0.9564	0.137917499533028\\
0.958	0.160903749455199\\
0.9602	0.183889999377371\\
0.961	0.206876249299542\\
0.9621	0.229862499221713\\
0.9633	0.252848749143885\\
0.9645	0.275834999066056\\
0.9662	0.298821248988228\\
0.9677	0.321807498910399\\
0.9694	0.34479374883257\\
0.9705	0.367779998754742\\
0.972	0.390766248676913\\
0.973	0.413752498599085\\
0.975	0.436738748521256\\
0.9765	0.459724998443427\\
0.9788	0.482711248365599\\
0.9799	0.50569749828777\\
0.9813	0.528683748209942\\
0.9824	0.551669998132113\\
0.9833	0.574656248054284\\
0.985	0.597642497976456\\
0.9858	0.620628747898627\\
0.9872	0.643614997820798\\
0.9876	0.66660124774297\\
0.9882	0.689587497665141\\
0.9888	0.712573747587313\\
0.99	0.735559997509484\\
0.9909	0.758546247431655\\
0.992	0.781532497353827\\
0.9924	0.804518747275998\\
0.9933	0.82750499719817\\
0.9943	0.850491247120341\\
0.9954	0.873477497042512\\
0.9966	0.896463746964684\\
0.9973	0.919449996886855\\
0.9982	0.942436246809026\\
0.9991	0.965422496731198\\
0.9996	0.988408746653369\\
0.9997	1.01139499657554\\
0.9997	1.03438124649771\\
0.9998	1.05736749641988\\
0.9999	1.08035374634205\\
1	1.10333999626423\\
};\addlegendentry{$1-\lambda=\text{0.35}, \epsu=\text{0.525}$}

\addplot [mark=none, color=mycolor1, line width=2.0pt] coordinates {(0.91,0.5) (1,0.5)};
\node (a) at (0.988,0.52) {(0.99,0.5)};

\addplot +[mark=none, color=mycolor2, line width=2.0pt] coordinates {(0.91,0.35) (1,0.35)};
\node  at (0.965,0.38) {(0.97,0.35)};
\end{axis}
\begin{axis}[%
	width=5.833in,
	height=4.375in,
	at={(0in,0in)},
	scale only axis,
	xmin=0,
	xmax=1,
	ymin=0,
	ymax=1,
	axis line style={draw=none},
	ticks=none,
	axis x line*=bottom,
	axis y line*=left
	]
\end{axis}

\end{tikzpicture}%
}}
		\caption{\label{fig:eps1} Privacy-utility trade-off for $\eps=\epsu+\epsl=1.5$ where  $\epsl=\lambda\eps$ and $\epsu=(1-\lambda)\eps$.} 
	\end{figure*}
	\begin{figure*}
		\centering     %%% not \center
		\subfigure[\label{fig:eps2 utility} Utility]{\scalebox{0.333}{\definecolor{mycolor1}{rgb}{0,0.447,0.741}%
\definecolor{mycolor2}{rgb}{0.85,0.325,0.098}%
\begin{tikzpicture}

\begin{axis}[%
width=6in,
height=3.71in,
at={(0.758in,0.516in)},
scale only axis,
xmin=0.3,
xmax=1,
xlabel style={font=\color{white!15!black}},
xlabel={\huge CDF},
xtick={0.3,0.4,0.5,0.6,0.7,0.8,0.9,1},
every x tick label/.append style={font=\color{darkgray!60!black},font=\huge},
ymin=0,
ymax=0.35,
ylabel style={font=\color{white!15!black}},
ytick={0,0.05,0.1,0.15,0.2,0.25,0.3,0.35},
ylabel={\huge Utility},
every y tick label/.append style={font=\color{darkgray!60!black},font=\Large},
axis background/.style={fill=white},
title style={font=\bfseries},
xmajorgrids,
ymajorgrids,
legend style={at={(0.01,0.82)}, anchor=south west, legend cell align=left, align=left, draw=white!15!black,font=\huge}
]
%\addplot [color=mycolor1, line width=3.0pt,mark=pentagon, mark options={solid},mark size=2pt]
%  table[row sep=crcr]{%
%0.9715	-4.44089209850063e-16\\
%0.9715	0.00210553776463316\\
%0.9715	0.00421107552926677\\
%0.9715	0.00631661329390037\\
%0.9715	0.00842215105853398\\
%0.9715	0.0105276888231676\\
%0.9715	0.0126332265878012\\
%0.9715	0.0147387643524348\\
%0.9715	0.0168443021170684\\
%0.9715	0.018949839881702\\
%0.9715	0.0210553776463356\\
%0.9715	0.0231609154109692\\
%0.9715	0.0252664531756028\\
%0.9715	0.0273719909402364\\
%0.9715	0.02947752870487\\
%0.9715	0.0315830664695037\\
%0.9715	0.0336886042341373\\
%0.9715	0.0357941419987709\\
%0.9715	0.0378996797634045\\
%0.9716	0.0400052175280381\\
%0.9729	0.0421107552926717\\
%0.9759	0.0442162930573053\\
%0.9791	0.0463218308219389\\
%0.9841	0.0484273685865725\\
%0.99	0.0505329063512061\\
%0.9951	0.0526384441158397\\
%0.9984	0.0547439818804733\\
%0.9996	0.0568495196451069\\
%0.9999	0.0589550574097405\\
%0.9999	0.0610605951743741\\
%0.9999	0.0631661329390077\\
%0.9999	0.0652716707036414\\
%0.9999	0.067377208468275\\
%0.9999	0.0694827462329086\\
%0.9999	0.0715882839975422\\
%0.9999	0.0736938217621758\\
%0.9999	0.0757993595268094\\
%0.9999	0.077904897291443\\
%0.9999	0.0800104350560766\\
%0.9999	0.0821159728207102\\
%0.9999	0.0842215105853438\\
%0.9999	0.0863270483499774\\
%0.9999	0.088432586114611\\
%0.9999	0.0905381238792446\\
%0.9999	0.0926436616438782\\
%0.9999	0.0947491994085118\\
%0.9999	0.0968547371731455\\
%0.9999	0.098960274937779\\
%1	0.101065812702413\\
%};
%\addlegendentry{$\lambda\text{=0.35}$}

\addplot [color=mycolor1, line width=3.0pt,mark=x,mark options={solid},mark size=3pt]
  table[row sep=crcr]{%
0.7216	-4.44089209850063e-16\\
0.7216	0.00398225496915452\\
0.7216	0.00796450993830948\\
0.7216	0.0119467649074644\\
0.7216	0.0159290198766194\\
0.7216	0.0199112748457744\\
0.7216	0.0238935298149293\\
0.7217	0.0278757847840843\\
0.7221	0.0318580397532392\\
0.7272	0.0358402947223942\\
0.7521	0.0398225496915492\\
0.823	0.0438048046607041\\
0.9061	0.0477870596298591\\
0.9506	0.051769314599014\\
0.9588	0.055751569568169\\
0.959	0.059733824537324\\
0.959	0.0637160795064789\\
0.959	0.0676983344756339\\
0.959	0.0716805894447888\\
0.9591	0.0756628444139438\\
0.9595	0.0796450993830988\\
0.9618	0.0836273543522537\\
0.9665	0.0876096093214087\\
0.9756	0.0915918642905636\\
0.9863	0.0955741192597186\\
0.9937	0.0995563742288735\\
0.9962	0.103538629198029\\
0.9967	0.107520884167183\\
0.9968	0.111503139136338\\
0.9968	0.115485394105493\\
0.9968	0.119467649074648\\
0.9968	0.123449904043803\\
0.9968	0.127432159012958\\
0.9971	0.131414413982113\\
0.9977	0.135396668951268\\
0.9982	0.139378923920423\\
0.9988	0.143361178889578\\
0.9991	0.147343433858733\\
0.9993	0.151325688827888\\
0.9996	0.155307943797043\\
0.9996	0.159290198766198\\
0.9996	0.163272453735353\\
0.9996	0.167254708704508\\
0.9996	0.171236963673663\\
0.9996	0.175219218642818\\
0.9996	0.179201473611973\\
0.9998	0.183183728581128\\
0.9998	0.187165983550283\\
1	0.191148238519438\\
};
\addlegendentry{$\lambda=\text{0.50}$}

\addplot [color=mycolor2, line width=3.0pt,mark=oplus,mark options={solid},mark size=3pt]
  table[row sep=crcr]{%
0.3163	-4.44089209850063e-16\\
0.3163	0.0065219551665417\\
0.3163	0.0130439103330838\\
0.3163	0.019565865499626\\
0.3163	0.0260878206661681\\
0.3185	0.0326097758327102\\
0.4072	0.0391317309992524\\
0.6376	0.0456536861657945\\
0.6938	0.0521756413323367\\
0.6945	0.0586975964988788\\
0.6945	0.0652195516654209\\
0.6948	0.0717415068319631\\
0.6986	0.0782634619985052\\
0.7331	0.0847854171650474\\
0.8232	0.0913073723315895\\
0.8871	0.0978293274981317\\
0.8964	0.104351282664674\\
0.8969	0.110873237831216\\
0.8971	0.117395192997758\\
0.8995	0.1239171481643\\
0.9131	0.130439103330842\\
0.9401	0.136961058497384\\
0.9654	0.143483013663927\\
0.9727	0.150004968830469\\
0.9733	0.156526923997011\\
0.9733	0.163048879163553\\
0.9743	0.169570834330095\\
0.9781	0.176092789496637\\
0.9861	0.182614744663179\\
0.9924	0.189136699829722\\
0.9945	0.195658654996264\\
0.995	0.202180610162806\\
0.995	0.208702565329348\\
0.9953	0.21522452049589\\
0.9962	0.221746475662432\\
0.9979	0.228268430828974\\
0.9989	0.234790385995517\\
0.9993	0.241312341162059\\
0.9993	0.247834296328601\\
0.9993	0.254356251495143\\
0.9993	0.260878206661685\\
0.9993	0.267400161828227\\
0.9994	0.273922116994769\\
0.9997	0.280444072161312\\
0.9999	0.286966027327854\\
0.9999	0.293487982494396\\
0.9999	0.300009937660938\\
0.9999	0.30653189282748\\
1	0.313053847994022\\
};
\addlegendentry{$\lambda=\text{0.65}$}

%\addplot [color=mycolor4, line width=3.0pt, mark=diamond,
%mark options={solid},
%mark size=2pt]
%  table[row sep=crcr]{%
%0.3657	-4.44089209850063e-16\\
%0.3657	0.00667919815199992\\
%0.3657	0.0133583963040003\\
%0.3657	0.0200375944560007\\
%0.3657	0.026716792608001\\
%0.3679	0.0333959907600014\\
%0.459	0.0400751889120017\\
%0.6969	0.0467543870640021\\
%0.7353	0.0534335852160025\\
%0.7356	0.0601127833680028\\
%0.7356	0.0667919815200032\\
%0.7356	0.0734711796720036\\
%0.7384	0.0801503778240039\\
%0.7722	0.0868295759760043\\
%0.865	0.0935087741280047\\
%0.9149	0.100187972280005\\
%0.9191	0.106867170432005\\
%0.9192	0.113546368584006\\
%0.9192	0.120225566736006\\
%0.9209	0.126904764888007\\
%0.9337	0.133583963040007\\
%0.9584	0.140263161192007\\
%0.9766	0.146942359344008\\
%0.9797	0.153621557496008\\
%0.9798	0.160300755648008\\
%0.9799	0.166979953800009\\
%0.9805	0.173659151952009\\
%0.9838	0.180338350104009\\
%0.9911	0.18701754825601\\
%0.9952	0.19369674640801\\
%0.9963	0.200375944560011\\
%0.9964	0.207055142712011\\
%0.9964	0.213734340864011\\
%0.9964	0.220413539016012\\
%0.9972	0.227092737168012\\
%0.9986	0.233771935320012\\
%0.9994	0.240451133472013\\
%0.9994	0.247130331624013\\
%0.9995	0.253809529776013\\
%0.9995	0.260488727928014\\
%0.9995	0.267167926080014\\
%0.9996	0.273847124232015\\
%0.9998	0.280526322384015\\
%0.9999	0.287205520536015\\
%0.9999	0.293884718688016\\
%0.9999	0.300563916840016\\
%0.9999	0.307243114992016\\
%0.9999	0.313922313144017\\
%1	0.320601511296017\\
%};
%\addlegendentry{$\lambda\text{=0.70}$}

\end{axis}

\begin{axis}[%
width=5.833in,
height=4.375in,
at={(0in,0in)},
scale only axis,
xmin=0,
xmax=1,
ymin=0,
ymax=1,
axis line style={draw=none},
ticks=none,
axis x line*=bottom,
axis y line*=left
]
\end{axis}
\end{tikzpicture}%
}}
		\subfigure[\label{fig:eps2 leakagel}Privacy leakage lowebound]{\scalebox{0.333}{\definecolor{mycolor1}{rgb}{0,0.447,0.741}%
\definecolor{mycolor2}{rgb}{0.85,0.325,0.098}%
\begin{tikzpicture}

\begin{axis}[%
width=6in,
height=3.7in,
at={(0.758in,0.522in)},
scale only axis,
xmin=0.5,
xmax=1,
xlabel style={font=\color{white!15!black}},
xlabel={\huge CDF},
every x tick label/.append style={font=\color{darkgray!60!black},font=\huge},
ymin=0,
ymax=12,
ylabel style={font=\color{white!15!black}},
ylabel={\huge $\max_{y}|\nu(y)|$},
ytick={0,1,2,3,4,5,6,7,8,9,10,11,12},
every y tick label/.append style={font=\color{darkgray!60!black},font=\huge},
axis background/.style={fill=white},
title style={font=\bfseries},
xmajorgrids,
ymajorgrids,
legend style={at={(0.01,0.82)}, anchor=south west, legend cell align=left, align=left, draw=white!15!black,font=\huge}
]
\addplot [color=mycolor1, line width=2.0pt,mark=x,mark options={solid},mark size=3pt]
  table[row sep=crcr]{%
0.8472	0\\
%0.8578	0.103022014676086\\
0.8674	0.206044029352172\\
%0.8774	0.309066044028259\\
0.8872	0.412088058704345\\
%0.897	0.515110073380431\\
0.9067	0.618132088056517\\
%0.9142	0.721154102732603\\
0.9222	0.82417611740869\\
%0.9288	0.927198132084776\\
0.9349	1.03022014676086\\
%0.9403	1.13324216143695\\
0.947	1.23626417611303\\
%0.9512	1.33928619078912\\
0.9561	1.44230820546521\\
%0.9593	1.54533022014129\\
0.9628	1.64835223481738\\
%0.9667	1.75137424949347\\
0.9691	1.85439626416955\\
%0.9725	1.95741827884564\\
0.9756	2.06044029352172\\
%0.9776	2.16346230819781\\
0.9796	2.2664843228739\\
%0.981	2.36950633754998\\
0.9833	2.47252835222607\\
%0.9856	2.57555036690216\\
0.9872	2.67857238157824\\
%0.9886	2.78159439625433\\
0.9895	2.88461641093041\\
%0.9904	2.9876384256065\\
0.9909	3.09066044028259\\
%0.9916	3.19368245495867\\
0.9929	3.29670446963476\\
%0.9934	3.39972648431084\\
0.9937	3.50274849898693\\
%0.9944	3.60577051366302\\
0.9948	3.7087925283391\\
%0.9954	3.81181454301519\\
0.9958	3.91483655769128\\
%0.9963	4.01785857236736\\
0.9966	4.12088058704345\\
%0.9973	4.22390260171953\\
0.9975	4.32692461639562\\
%0.9975	4.42994663107171\\
0.9976	4.53296864574779\\
%0.9978	4.63599066042388\\
0.9979	4.73901267509997\\
%0.9983	4.84203468977605\\
0.9983	4.94505670445214\\
%0.9986	5.04807871912822\\
0.9986	5.15110073380431\\
%0.9987	5.2541227484804\\
0.9989	5.35714476315648\\
%0.9989	5.46016677783257\\
0.9992	5.56318879250866\\
%0.9992	5.66621080718474\\
0.9993	5.76923282186083\\
%0.9993	5.87225483653691\\
0.9993	5.975276851213\\
%0.999499999999999	6.07829886588909\\
0.999499999999999	6.18132088056517\\
%0.999499999999999	6.28434289524126\\
0.999499999999999	6.38736490991734\\
%0.999499999999999	6.49038692459343\\
0.999499999999999	6.59340893926952\\
%0.999499999999999	6.6964309539456\\
0.999499999999999	6.79945296862169\\
%0.999499999999999	6.90247498329778\\
0.999499999999999	7.00549699797386\\
%0.999499999999999	7.10851901264995\\
0.999499999999999	7.21154102732603\\
%0.999499999999999	7.31456304200212\\
0.999599999999999	7.41758505667821\\
%0.999599999999999	7.52060707135429\\
0.999599999999999	7.62362908603038\\
%0.999699999999999	7.72665110070646\\
0.999799999999999	7.82967311538255\\
%0.999799999999999	7.93269513005864\\
0.999799999999999	8.03571714473472\\
%0.999799999999999	8.13873915941081\\
0.999799999999999	8.2417611740869\\
%0.999799999999999	8.34478318876298\\
0.999799999999999	8.44780520343907\\
%0.999799999999999	8.55082721811515\\
0.999899999999999	8.65384923279124\\
%0.999899999999999	8.75687124746733\\
0.999899999999999	8.85989326214341\\
%0.999899999999999	8.9629152768195\\
0.999899999999999	9.06593729149558\\
%0.999899999999999	9.16895930617167\\
0.999899999999999	9.27198132084776\\
%0.999899999999999	9.37500333552384\\
0.999899999999999	9.47802535019993\\
%0.999899999999999	9.58104736487602\\
0.999899999999999	9.6840693795521\\
%0.999899999999999	9.78709139422819\\
0.999899999999999	9.89011340890428\\
%0.999899999999999	9.99313542358036\\
0.999999999999999	10.0961574382564\\
};
\addlegendentry{$\lambda=\text{0.50, }\epsl \text{=1}$}

\addplot [color=mycolor2, line width=2.0pt,mark=oplus,mark options={solid},mark size=3pt]
  table[row sep=crcr]{%
0.5663	0\\
%0.5957	0.115582642886227\\
0.6233	0.231165285772454\\
%0.6525	0.346747928658681\\
0.6804	0.462330571544909\\
%0.7069	0.577913214431136\\
0.7309	0.693495857317363\\
%0.7563	0.80907850020359\\
0.7781	0.924661143089817\\
%0.7964	1.04024378597604\\
0.8153	1.15582642886227\\
%0.8341	1.2714090717485\\
0.8496	1.38699171463473\\
%0.8642	1.50257435752095\\
0.8773	1.61815700040718\\
%0.8903	1.73373964329341\\
0.9003	1.84932228617963\\
%0.9089	1.96490492906586\\
0.9179	2.08048757195209\\
%0.928	2.19607021483832\\
0.9347	2.31165285772454\\
%0.9417	2.42723550061077\\
0.949	2.542818143497\\
%0.9543	2.65840078638322\\
0.9583	2.77398342926945\\
%0.9624	2.88956607215568\\
0.9657	3.00514871504191\\
%0.9681	3.12073135792813\\
0.9717	3.23631400081436\\
%0.9742	3.35189664370059\\
0.9766	3.46747928658681\\
%0.9795	3.58306192947304\\
0.9814	3.69864457235927\\
%0.9835	3.8142272152455\\
0.985	3.92980985813172\\
%0.9867	4.04539250101795\\
0.9882	4.16097514390418\\
%0.9896	4.2765577867904\\
0.9903	4.39214042967663\\
%0.991	4.50772307256286\\
0.9918	4.62330571544909\\
%0.9927	4.73888835833531\\
0.9936	4.85447100122154\\
%0.9938	4.97005364410777\\
0.9946	5.08563628699399\\
%0.9951	5.20121892988022\\
0.9954	5.31680157276645\\
%0.9957	5.43238421565268\\
0.9963	5.5479668585389\\
%0.9965	5.66354950142513\\
0.9967	5.77913214431136\\
%0.9971	5.89471478719758\\
0.9975	6.01029743008381\\
%0.9977	6.12588007297004\\
0.9977	6.24146271585627\\
%0.998	6.35704535874249\\
0.9982	6.47262800162872\\
%0.9983	6.58821064451495\\
0.9984	6.70379328740117\\
%0.9986	6.8193759302874\\
0.9986	6.93495857317363\\
%0.9987	7.05054121605986\\
0.9987	7.16612385894608\\
%0.9988	7.28170650183231\\
0.9989	7.39728914471854\\
%0.9989	7.51287178760476\\
0.999	7.62845443049099\\
%0.9991	7.74403707337722\\
0.9992	7.85961971626344\\
%0.9993	7.97520235914967\\
0.9994	8.0907850020359\\
%0.9994	8.20636764492213\\
0.999499999999999	8.32195028780835\\
%0.999599999999999	8.43753293069458\\
0.999599999999999	8.55311557358081\\
%0.999699999999999	8.66869821646704\\
0.999699999999999	8.78428085935326\\
%0.999699999999999	8.89986350223949\\
0.999699999999999	9.01544614512572\\
%0.999699999999999	9.13102878801194\\
0.999699999999999	9.24661143089817\\
%0.999699999999999	9.3621940737844\\
0.999699999999999	9.47777671667062\\
%0.999699999999999	9.59335935955685\\
0.999699999999999	9.70894200244308\\
%0.999699999999999	9.82452464532931\\
0.999699999999999	9.94010728821553\\
%0.999699999999999	10.0556899311018\\
0.999799999999999	10.171272573988\\
%0.999799999999999	10.2868552168742\\
0.999799999999999	10.4024378597604\\
%0.999799999999999	10.5180205026467\\
0.999799999999999	10.6336031455329\\
%0.999799999999999	10.7491857884191\\
0.999799999999999	10.8647684313054\\
%0.999799999999999	10.9803510741916\\
0.999799999999999	11.0959337170778\\
%0.999899999999999	11.211516359964\\
0.999999999999999	11.3270990028503\\
};
\addlegendentry{$\lambda=\text{0.65, }\epsl\text{=1.3}$}

\addplot [mark=none, color=mycolor2, line width=2.0pt] coordinates {(0.5,1.3) (1,1.3)};
\node (a) at (0.825,1.55) {(0.84,1.3)};

\end{axis}

\begin{axis}[%
width=5.833in,
height=4.375in,
at={(0in,0in)},
scale only axis,
xmin=0,
xmax=1,
ymin=0,
ymax=1,
axis line style={draw=none},
ticks=none,
axis x line*=bottom,
axis y line*=left
]
\end{axis}
\end{tikzpicture}%
}}
		\subfigure[\label{fig:eps2 leakageu}Privacy lekage upperbound]{\scalebox{0.333}{\definecolor{mycolor1}{rgb}{0,0.447,0.741}%
\definecolor{mycolor2}{rgb}{0.85,0.325,0.098}%
\begin{tikzpicture}

\begin{axis}[%
width=6in,
height=3.7in,
at={(0.758in,0.522in)},
scale only axis,
xmin=0.3,
xmax=1,
xlabel style={font=\color{white!15!black}, font=\large},
xtick={0.3,0.4,0.5,0.6,0.7,0.8,0.9,1},
every x tick label/.append style={font=\color{darkgray!60!black},font=\huge},
xlabel={\huge CDF},
ymin=0,
ymax=1.4,
ylabel style={font=\color{white!15!black}},
every y tick label/.append style={font=\color{darkgray!60!black},font=\huge},
ylabel={\huge $\max_{y}\xi(y)$},
axis background/.style={fill=white},
xmajorgrids,
ymajorgrids,
legend style={at={(0.01,0.82)}, anchor=south west, legend cell align=left, align=left, draw=white!15!black,font=\huge}
]
\addplot [color=mycolor1, line width=2.0pt,mark=x,mark options={solid},mark size=3pt]
  table[row sep=crcr]{%
0.7216	-6.66133814775094e-16\\
%0.8134	0.0119669550364394\\
0.8641	0.0239339100728795\\
%0.8719	0.0359008651093196\\
0.8739	0.0478678201457596\\
%0.8751	0.0598347751821997\\
0.8765	0.0718017302186398\\
%0.8788	0.0837686852550799\\
0.8801	0.0957356402915199\\
%0.8809	0.10770259532796\\
0.8823	0.1196695503644\\
%0.8834	0.13163650540084\\
0.8855	0.14360346043728\\
%0.8871	0.15557041547372\\
0.8886	0.16753737051016\\
%0.891	0.1795043255466\\
0.8929	0.191471280583041\\
%0.8936	0.203438235619481\\
0.8958	0.215405190655921\\
%0.8977	0.227372145692361\\
0.9	0.239339100728801\\
%0.9017	0.251306055765241\\
0.9049	0.263273010801681\\
%0.9066	0.275239965838121\\
0.9079	0.287206920874561\\
%0.9093	0.299173875911001\\
0.9111	0.311140830947441\\
%0.9136	0.323107785983881\\
0.9148	0.335074741020321\\
%0.9171	0.347041696056762\\
0.9183	0.359008651093202\\
%0.9202	0.370975606129642\\
0.9218	0.382942561166082\\
%0.924	0.394909516202522\\
0.9261	0.406876471238962\\
%0.9281	0.418843426275402\\
0.9298	0.430810381311842\\
%0.9314	0.442777336348282\\
0.9333	0.454744291384722\\
%0.9357	0.466711246421162\\
0.9373	0.478678201457602\\
%0.939	0.490645156494042\\
0.9411	0.502612111530483\\
%0.9427	0.514579066566923\\
0.9449	0.526546021603363\\
%0.9462	0.538512976639803\\
0.948	0.550479931676243\\
%0.9496	0.562446886712683\\
0.9517	0.574413841749123\\
%0.9534	0.586380796785563\\
0.9548	0.598347751822003\\
%0.9562	0.610314706858443\\
0.9574	0.622281661894883\\
%0.9598	0.634248616931323\\
0.9615	0.646215571967763\\
%0.9637	0.658182527004204\\
0.9654	0.670149482040644\\
%0.9671	0.682116437077084\\
0.9688	0.694083392113524\\
%0.9703	0.706050347149964\\
0.972	0.718017302186404\\
%0.9735	0.729984257222844\\
0.9751	0.741951212259284\\
%0.9763	0.753918167295724\\
0.9771	0.765885122332164\\
%0.9778	0.777852077368604\\
0.9791	0.789819032405044\\
%0.9808	0.801785987441485\\
0.9821	0.813752942477925\\
%0.9831	0.825719897514365\\
0.9841	0.837686852550805\\
%0.985	0.849653807587245\\
0.9861	0.861620762623685\\
%0.9872	0.873587717660125\\
0.9886	0.885554672696565\\
%0.9898	0.897521627733005\\
0.9912	0.909488582769445\\
%0.9923	0.921455537805885\\
0.9935	0.933422492842325\\
%0.9942	0.945389447878765\\
0.9947	0.957356402915206\\
%0.9958	0.969323357951646\\
0.9969	0.981290312988086\\
%0.9971	0.993257268024526\\
0.9975	1.00522422306097\\
%0.9977	1.01719117809741\\
0.9982	1.02915813313385\\
%0.998499999999999	1.04112508817029\\
0.998799999999999	1.05309204320673\\
%0.998999999999999	1.06505899824317\\
0.999199999999999	1.07702595327961\\
%0.999399999999999	1.08899290831605\\
0.999399999999999	1.10095986335249\\
%0.999699999999999	1.11292681838893\\
0.999699999999999	1.12489377342537\\
%0.999699999999999	1.13686072846181\\
0.999799999999999	1.14882768349825\\
%0.999899999999999	1.16079463853469\\
0.999999999999999	1.17276159357113\\
};\addlegendentry{$1-\lambda=\text{0.50, }\epsu=\text{1}$}

\addplot [color=mycolor2, line width=2.0pt,mark=oplus,mark options={solid},mark size=3pt]
  table[row sep=crcr]{%
0.3163	-4.44089209850063e-16\\
%0.4116	0.0125992744733509\\
0.5484	0.0251985489467022\\
%0.5946	0.0377978234200535\\
0.6136	0.0503970978934048\\
%0.6213	0.0629963723667561\\
0.6266	0.0755956468401074\\
%0.6315	0.0881949213134587\\
0.6357	0.10079419578681\\
%0.6386	0.113393470260161\\
0.6425	0.125992744733513\\
%0.6474	0.138592019206864\\
0.6524	0.151191293680215\\
%0.6572	0.163790568153567\\
0.6618	0.176389842626918\\
%0.6664	0.188989117100269\\
0.6699	0.201588391573621\\
%0.6738	0.214187666046972\\
0.6787	0.226786940520323\\
%0.6836	0.239386214993674\\
0.6889	0.251985489467026\\
%0.6943	0.264584763940377\\
0.6994	0.277184038413728\\
%0.7033	0.28978331288708\\
0.7084	0.302382587360431\\
%0.7133	0.314981861833782\\
0.72	0.327581136307134\\
%0.7258	0.340180410780485\\
0.7307	0.352779685253836\\
%0.7368	0.365378959727188\\
0.7418	0.377978234200539\\
%0.7483	0.39057750867389\\
0.7545	0.403176783147242\\
%0.7609	0.415776057620593\\
0.7666	0.428375332093944\\
%0.7716	0.440974606567295\\
0.7774	0.453573881040647\\
%0.7847	0.466173155513998\\
0.7896	0.478772429987349\\
%0.7952	0.491371704460701\\
0.8016	0.503970978934052\\
%0.8077	0.516570253407403\\
0.8142	0.529169527880755\\
%0.8201	0.541768802354106\\
0.8255	0.554368076827457\\
%0.8305	0.566967351300809\\
0.8365	0.57956662577416\\
%0.8431	0.592165900247511\\
0.8487	0.604765174720863\\
%0.854	0.617364449194214\\
0.8595	0.629963723667565\\
%0.8658	0.642562998140916\\
0.8715	0.655162272614268\\
%0.8765	0.667761547087619\\
0.8823	0.68036082156097\\
%0.8874	0.692960096034322\\
0.8934	0.705559370507673\\
%0.8994	0.718158644981024\\
0.9042	0.730757919454376\\
%0.909	0.743357193927727\\
0.9133	0.755956468401078\\
%0.9173	0.76855574287443\\
0.921	0.781155017347781\\
%0.9257	0.793754291821132\\
0.9294	0.806353566294484\\
%0.934	0.818952840767835\\
0.9378	0.831552115241186\\
%0.9422	0.844151389714538\\
0.9467	0.856750664187889\\
%0.9526	0.86934993866124\\
0.9575	0.881949213134591\\
%0.9628	0.894548487607943\\
0.9681	0.907147762081294\\
%0.9725	0.919747036554645\\
0.9767	0.932346311027997\\
%0.9805	0.944945585501348\\
0.9838	0.957544859974699\\
%0.987	0.970144134448051\\
0.9891	0.982743408921402\\
%0.9905	0.995342683394753\\
0.9923	1.0079419578681\\
%0.9933	1.02054123234146\\
0.9945	1.03314050681481\\
%0.9958	1.04573978128816\\
0.9966	1.05833905576151\\
0.9973	1.07093833023486\\
%0.998	1.08353760470821\\
0.9983	1.09613687918156\\
%0.9985	1.10873615365491\\
0.9989	1.12133542812827\\
%0.9992	1.13393470260162\\
0.9994	1.14653397707497\\
%0.9997	1.15913325154832\\
0.9997	1.17173252602167\\
%0.9999	1.18433180049502\\
0.9999	1.19693107496837\\
%0.9999	1.20953034944173\\
0.9999	1.22212962391508\\
1	1.23472889838843\\
};
\addlegendentry{$1-\lambda=\text{0.65, }\epsu=\text{0.7}$}

\addplot [mark=none,color=mycolor2, line width=2.0pt] coordinates {(0.3,0.7) (1,0.7)};
\node  at (0.84,0.73) {(0.89,0.7)};

\end{axis}

\begin{axis}[%
width=5.833in,
height=4.375in,
at={(0in,0in)},
scale only axis,
xmin=0,
xmax=1,
ymin=0,
ymax=1,
axis line style={draw=none},
ticks=none,
axis x line*=bottom,
axis y line*=left
]
\end{axis}
\end{tikzpicture}%
}}
		\caption{\label{fig:eps2} Privacy-utility trade-off for $\eps=\epsu+\epsl=2$ where  $\epsl=\lambda\eps$ and $\epsu=(1-\lambda)\eps$.}
	\end{figure*}
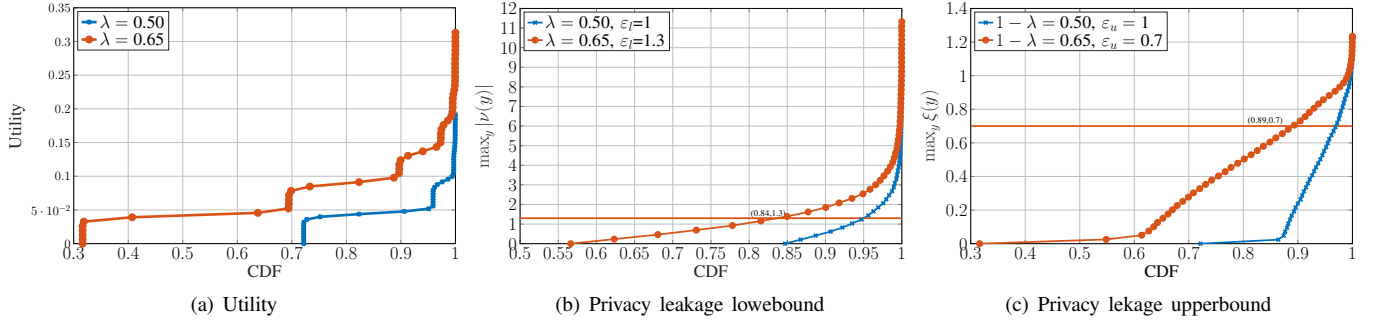
	
\subsection{ALIP Privacy-Utility trade-off and LDP}
	Proposition \refeq{prop:LDP}-\ref{prop:ALIP properties}) explains the relationship between LDP and ALIP, implying that achieving  $(\epsl,\epsu)$-ALIP guarantees $\eps$-LDP where $\eps=\epsl+\epsu$.
	We introduce $\lambda$ interpretation for a fixed value of $\eps$ to have different ALIP scenarios.
	 Here, $\epsl=\lambda\eps$ and $\epsu=(1-\lambda)\eps$ for $\lambda \in [0,1]$, and SLIP is given by $\lambda=0.5$ where $\epsl=\epsu=\frac{\eps}{2}.$
	In this regard, Figs.~\ref{fig:eps1} and \ref{fig:eps2}  illustrate the cumulative distribution function (CDF) of the utility and privacy leakage for the $10^4$ generated $\PSX$ where ${\eps}\in\{1.5,2\}$. 

	When $\lambda < 0.5$, the ALIP utility is smaller than the SLIP utility due to significantly restricted constraint  on the min-lift ($\nu(x) <\frac{\eps}{2}$), so we ignore such cases.
	When $\lambda>0.5$, ALIP enhances utility compared to SLIP by relaxation of min-lift ($\nu(x) > \frac{\eps}{2}$).
	In comparison of ALIP with SLIP ($\lambda=0.5$), we have shown the value of $\lambda$ results in the best overall utility enhancement in our numerical simulations among $\lambda \in (0.5, 1]$, which is $\lambda=0.65$.

	For $\eps=1.5$, Fig.~\ref{fig:eps1utility} shows the utility enhancement.
	In this case, the utility increment is not much due to the very restricted total LDP privacy budget $\eps=1.5$. 
	Fig.~\ref{fig:eps2 utility} demonstrates more utility enhancement between $\lambda=0.65$ and $\lambda=0.5$ for $\eps=2$ compared to $\eps=1.5$. 
	For example in Fig. ~\ref{fig:eps2 utility}, when $\lambda = 0.5$, only about $30\%$ of cases, have some NMI as utility, but when $\lambda = 0.65$, $70\%$ of cases show non-zero NMI as utility; while These percentages are $11\%$ and $4\%$ in Fig.~\ref{fig:eps1utility}.
	For both $\eps$ values utility have been increased in all $10^4$ generated distributions. 
	 
	Although complete merging minimizes privacy leakage in $\X_H$, even in SLIP, there is a chance that after watchdog randomization, the privacy budget in $X_H$ is not attained \cite{2020SadeghiITW,2020obfuscationlift}.
	 We study this for the asymmetric case here.
	 Figs.~\ref{fig:eps1leakagel} and \ref{fig:eps1leakagel} show the CDF of the privacy leakage for $\eps=1.5$, after randomization. 
	 In this case, $98.5\%$ and $96.8\%$ of distributions achieve the lower bound of privacy constraints $\epsl$, for $\lambda=0.5$ and $\lambda=0.65$ respectively. 
	 These values are $99\%$ and $97\%$ for the upper bound leakage. 
	 In Figs \ref{fig:eps2 leakagel} and \ref{fig:eps2 leakageu}, these percentages decrease to $93\%$ and $84\%$ for the lower bound and $98\%$ and $89\%$ for the upper bound leakage when $\eps=2$. 
	 The reduction in the percentage of cases attaining privacy happens since $X_H$'s size decreases due to relaxation of the privacy, which makes attaining privacy constraints more challenging. 
	 If the privacy constraints are not satisfied, a simple idea is to move more elements form $X_L$ to $X_H$ (at the cost of
	 reduced utility). 
%%%%%%%%%%%%%%%%%%%%%%%%%%%%%%%%%%%%%%%%%%%%%%%%%%%%%%%%%%%%%%%%%%%%%%%%%%%%%%%%%%%%%%%%%%%%%%%%%%%%%%%%%%%%%%%%%%%%%%%%%%%%%%%%%%%%%%%%%%%%%%%%%%%%%%%%%%%%%%%%%%%%%%%%%%%%%%%%%%
	
	\section{Conclusion}
	In this paper, we proposed a generalized definition of LIP and applied it to the watchdog mechanism where different values of privacy budgets are allocated to the maximum and minimum of log-lift. 
	We called it asymmetric local information privacy.
	Then we investigated the privacy-utility trade-off in this mechanism.
	It is demonstrated that for a fixed privacy budget on LDP, ALIP can enhance utility when we have relaxation on minimum lift values while restricting maximum lift values.
	Moreover, since other privacy measures such as MI, Sibson MI, $\alpha$-lift, $\alpha$-leakage are upper bounded by the maximum lift, ALIP tightly bounds these measures compared with symmetric-LIP.
	
	For future works, it is worth considering other privacy mechanisms rather than the watchdog mechanism by investigating the effects of ALIP on the privacy-utility trade-off for them. 
	Estimation of the distribution of log-lift is also an open problem that could be considered in connection with ALIP. 
	Combination of other relaxation methods such as $(\eps,\delta)$  with ALIP can also be considered.	
%%%%%%%%%%%%%%%%%%%%%%%%%%%%%%%%%%%%%%%%%%%%%%%%%%%%%%%%%%%%%%%%%%%%%%%%%%%%%%%%%%%%%%%%%%%%%%%%%%%%%%%%%%%%%%%%%%%%%%%%%%%%%%%%%%%%%%%%%%%%%%%%%%%%%%%%%%%%%%%%%%%%%%%%%%%%%%%%%%
\appendix
\small
%%%%%%%%%%%%%%%%%%%%%%%%%%%%%%%%%%%%%%%%%%%%%%%%%%%%%%%%%%%%%%%%%%%%%%%%%%%%%%%%%%%%%%%%%%%%%%%%%%%%%%%%%%%%%%%%%%%%%%%%%%%
\subsection{Definitions}\label{app:definitions}

	\begin{definition}
		For discrete $(S,Y) \sim \PSY$ and $\alpha\in(1,\infty)$, the Arimoto MI which is equivalent to the $\alpha$-leakage \cite{2018_tunable_measureinformationleakage} is given by
		\begin{equation}\label{eq:arimiot mutual info}
				I_{\alpha}^{A}(S;Y) \triangleq \frac{\alpha}{\alpha-1} \log \frac{\mathbb{E}_{Y}\left[\left\|P_{S|Y}(\cdot| 	Y)\right\|_{\alpha}\right]}{\left\|P_{S}\right\|_{\alpha}} .
		\end{equation}	
	\end{definition}
	
	\begin{definition}	
		For discrete $Y \sim \PY$ with alphabet $\Y$, a distribution $Q_Y$ over $\Y$, and $\alpha\in(1,\infty)$, the Rényi divergence is given by
		\begin{equation}\label{eq:Renyi divergence}
			D_{\alpha}(\PY||Q_Y) \triangleq \frac{1}{\alpha-1} \log \left(\sum_{y \in \Y} \PY(y)^{\alpha} Q_Y(y)^{1-\alpha}\right).
		\end{equation}
		Then, for discrete $(S,Y) \sim \PSY$, the Sibson's MI is given by
		\begin{equation}
			\begin{aligned}
				I_{\alpha}^{S}(S;Y) & \triangleq \inf _{Q_{Y}} D_{\alpha}\left(\PSY \| \PS \times Q_{Y}\right) \\
				&=\frac{\alpha}{\alpha-1} \log \sum_{y \in \Y}\left(\sum_{s \in \Sen} \PS(s) P_{Y|S}(y|s)^{\alpha}\right)^{\frac{1}{\alpha}}.
			\end{aligned}
		\end{equation}
	\end{definition}

	\begin{definition}
		For discrete $(S,Y) \sim \PSY$, and $\alpha \in (1,\infty)$, the $\alpha$-lift \cite{2021alphading} is given by
		\begin{equation} \label{eq:AlphaLiftX}
			\ell_{\alpha}(y) \triangleq  \left(\sum_{s \in \Sen}\PS(s)\left(\frac{\PSY(s,y)}{\PS(s) \PY(y)}\right)^{\alpha}\right)^{1 /\alpha}, \quad \forall y \in \Y.
		\end{equation}
		Then, the Sibson's MI will be
		\begin{equation}
			I_{\alpha}^{S}(S;Y) = \frac{\alpha}{\alpha-1} \log \E_{Y}[\ell_{\alpha}(Y)] , \quad \forall \alpha \in (1,\infty).
		\end{equation}
	\end{definition}

	For $\alpha=1$, Sibson and Arimoto MI reduce to Shannon's MI. For $\alpha=\infty$, the Arimoto MI will be
	\begin{equation}
		I_{\infty}^{A}(S;Y)=\log \frac{\sum_{y} \max_{s} P_{SY}(s,y)}{\max_{s} P_{S}(s)}.
	\end{equation}
	$\alpha$-lift reduces to maximum of the lift $\ell_{\infty}(y)=\max_{s \in \Sen}\ell(s,y)$, and the Sibson MI is given by 
	\begin{equation}
		I_{\infty}^{S}(S;Y)= \log \E_{Y}[\max_{s \in \Sen}\ell(s,Y)],
	\end{equation}
	which is equivalent to the \textit{maximal leakage}.
%%%%%%%%%%%%%%%%%%%%%%%%%%%%%%%%%%%%%%%%%%%%%%%%%%%%%%%%%%%%%%%%%%%%%%%%%%%%%%%%%%%%%%%%%%%%%%%%%%%%%%%%%%%%%%%%%%%%%%%%%%%%%%%%%%%%%%%%%%%%%%%%%%%%%%%%%%%%%%%%%%%
\subsection{Proof of Proposition \ref{prop:ALIP properties}}\label{app:proof of prop 1}
    Proposition \ref{prop:ALIP properties}-\ref{prop:LDP}: For any $s, s^{\prime} \in \mathcal{S}$, assume $\PSY(s,y)>0$ and $\PSY(s',y)>0$, by Definition \ref{Def:ALIP} we have:
   	\begin{align}
   		\nonumber \left|\log \frac{\PYgS(y|s)}{\PYgS(y|s^{'})}\right| 
   		&=\left|\log \frac{\PYgS(y|s)}{\PS(s)}-\log \frac{\PYgS(y|s^{'})}{\PS(s)}\right|
   		\leq  \eps_u+\eps_l
   	\end{align}
     Proposition \ref{prop:ALIP properties}-\ref{prop:MI}:  For mutual information we have:
    \begin{align}
    	\nonumber I(S;Y) & = \sum_{s,y} \PSY(s,y) \log \frac{\PSgY(s|y)}{\PS(s)} \\
    		   			 &\leq \sum_{s,y} \PSY(s,y) \epsu =\epsu.
    \end{align}
    For $\alpha$-lift, using Jensen inequality and Definition \ref{Def:ALIP}  we have: 
    \begin{equation}
        1\leq \ell_{\alpha}(y) \leq e^{\epsu}.
    \end{equation}
    For maximal leakage, note that:
    \begin{equation}
        0\leq \max_{s}\ell(s,y) \leq e^{\epsu} \Rightarrow 0\leq \log \E_{Y}[\max_{s\in\Sen}\ell(s,Y)] \leq \epsu
    \end{equation}
%%%%%%%%%%%%%%%%%%%%%%%%%%%%%%%%%%%%%%%%%%%%%%%%%%%%%%%%%%%%%%%%%%%%%%%%%%%%%%%%%%%%%%%%%%%%%%%%%%%%%%%%%
    Proposition \ref{prop:ALIP properties}-\ref{prop:Sibson}: 
    Definition \ref{Def:ALIP}, we have $\PSY(s,y)^{\alpha} \leq e^{\alpha \epsu}\PS(s)^{\alpha} \PY(y)^{\alpha}$, and
	\begin{equation}
	 	\begin{aligned}
			&D_{\alpha}\big(\PSY|\PS Q_{Y}\big) \\
			=& \frac{1}{\alpha-1} \log \left(\sum_{s,y} \frac{\PSY(s,y)^{\alpha}}{\PS(s)^{\alpha-1} Q_{Y}(y)^{\alpha-1}}\right) \\
			\leq& \frac{1}{\alpha-1} \log \left(e^{\alpha \epsu} \sum_{s,y} \frac{\PS(s)^{\alpha} \PY(y)^{\alpha}}{\PS(s)^{\alpha-1} Q_{Y}(y)^{\alpha-1}}\right) \\
			=& \frac{\alpha \epsu}{\alpha-1}+\frac{1}{\alpha-1} \log \left(\sum_{y} \frac{\PY(y)^{\alpha}}{Q_{Y}(y)^{\alpha-1}}\right) \\
			=& \frac{\alpha}{\alpha-1} \epsu+D_{\alpha}\left(\PY\| Q_{Y}\right) .
	 	\end{aligned}
	\end{equation}
  	 Therefore, since $\inf _{Q_{Y}} D_{\alpha}\left(\PY|Q_{Y}\right)=0$ when $\PY=Q_{Y}$, we have
    \begin{equation}
    	\begin{aligned}
    		I_{\alpha}^{S}(S;Y) & \leq \frac{\alpha}{\alpha-1} \epsu+\inf _{Q_{Y}(y)} D_{\alpha}\left(\PY|Q_{Y}\right) 
    							=\frac{\alpha}{\alpha-1} \epsu.
    	\end{aligned}
    \end{equation}
    For Arimoto MI, by the Definition \ref{Def:ALIP}, we have $P_{S|Y}(s|y)^{\alpha} \leq e^{\alpha \varepsilon_u} P_{S}(s)^{\alpha} \Rightarrow\left\|P_{S|Y}(\cdot|Y)\right\|_{\alpha} \leq e^{\varepsilon_u}\left\|P_{S}\right\|_{\alpha}.$ 
    Therefore,
	\begin{equation}
		\begin{aligned}
			I_{\alpha}^{A}(S;Y) &=\frac{\alpha}{\alpha-1} \log \frac{\sum_{y}\left\|\PSgY(\cdot|y)\right\|_{\alpha} P_{Y}(y)}{\left\|P_{S}\right\|_{\alpha}}\\
			& \leq \frac{\alpha}{\alpha-1} \log \frac{\sum_{y} e^{\varepsilon_u}\left\|P_{S}\right\|_{\alpha} P_{Y}(y)}{\left\|P_{S}\right\|_{\alpha}} 
			=\frac{\alpha}{\alpha-1} \varepsilon_u .
		\end{aligned}
	\end{equation}
%%%%%%%%%%%%%%%%%%%%%%%%%%%%%%%%%%%%%%%%%%%%%%%%%%%%%%%%%%%%%%%%%%%%%%%%%%%%%%%%%%%%%%%%%%%%%%%%%%%%%%%%%%%%%%%%%%%%%%%%%%%%%%%%%%%%%%%%%%%%%%%%%%%%%%%%%%%%%%%%%%%%
\subsection{Proof of Proposition \ref{prop:Xinvarinat}}\label{app:proof of prop 2}
	We follow the steps in proof of \cite[Corollary 2]{2020SadeghiITW}.
	Let $s_u\in\argmax\PXgS(\X_H|s)$ and $s_\ell\in\argmin\PXgS(\X_H|s)$. 
	Proposition \ref{prop:Xinvarinat} is proven by contradiction. 
	Assume there exists another $r_{Y|X}(y|x)$ that attains lifts strictly smaller than $e^{-\epsl^{*}}$ and $e^{\epsu^{*}}$. Then both:
	\begin{equation}\label{eq:ineq max}
		\frac{\displaystyle\sum_{x\in\X_H}r_{Y|X}(y|x)\PXgS(x|s)}{\displaystyle\sum_{x\in\X_H}r_{Y|X}(y|x)\PX(x)} < 	\frac{P\left(\X_H|s_u\right)}{P\left(\X_H\right)}:=\ell\left(s_u,\X_H\right)
	\end{equation}
	\begin{equation}\label{eq:ineq min}
		\frac{\displaystyle\sum_{x\in\X_H}r_{Y|X}(y|x)\PXgS(x|s)}{\displaystyle\sum_{x\in\X_H}r_{Y|X}(y|x)\PX(x)} > 	\frac{P\left(\X_H|s_\ell\right)}{P\left(\X_H\right)}:=\ell\left(s_\ell,\X_H\right)
	\end{equation}
	hold for all $y \in \X_H$ and $s \in \Sen$. However, for any $y$, we have
	\begin{align}
		\scriptstyle
		\nonumber&\frac{\sum_{x\in\X_H}r_{Y|X}(y|x)\PXgS(x|s)}{\sum_{x\in\X_H}r_{Y|X}(y|x)\PX(x)}- \frac{ P\left(\X_H|s_u\right)}{P\left(\X_H\right)}\\
		%\nonumber=&\frac{P\left(\X_H\right)\sum_{x\in\X_H}r_{Y|X}(y|x)\PXgS(x|s)}{P\left(\X_H\right)\sum_{x\in\X_H} r_{Y|X}(y|x)\PX(x)}\\
		%\nonumber&-\frac{P\left(\X_H|s_u\right)\sum_{x\in\X_H}r_{Y|X}(y|x)\PX(x)}{P\left(\X_H\right)\sum_{x\in\X_H} r_{Y|X}(y|x)\PX(x)}\\
		=& \frac{\sum_{x\in\X_H} r_{Y|X}(y|x)\left(P\left(\X_H\right)\PXgS(x|s)-P\left(\X_H|s_u\right) 	\PX(x)\right)}{P\left(\X_H\right)\sum_{x \in\X_H}r_{Y|X}(y|x)\PX(x)} \label{eq:17}\\
		\nonumber=& \frac{\sum_{x \in\X_H}\left(1-\sum_{y^{\prime} \in\X_H: y^{\prime} \neq y} r_{Y|X}\left(y^{\prime}|x\right)\right) }{P\left(\X_H\right) \sum_{x\in\X_H} r_{Y|X}(y|x)\PX(x)} \\
		&\times\frac{ \left(P(\X_H)\PXgS(x|s)-P\left(\X_H|s_u\right)\PX(x)\right)}{P\left(\X_H\right)\sum_{x \in\X_H}r_{Y|X}(y|x) \PX(x)}\\
		>& \frac{\sum_{x\in\X_H}\left(P\left(\X_H\right)\PXgS(x|s)-P\left(\X_H|s_u\right)\PX(x)\right)}{P\left(\X_H\right) \sum_{x\in\X_H}r_{Y|X}(y|x)\PX(x)} \label{eq:ineq proof}\\
		=& \frac{P\left(\X_H|s\right)-P\left(\X_H|s_u\right)}{\sum_{x\in\X_H}r_{Y|X}(y|x)\PX(x)} \label{eq:last proof}
	\end{align}
	\normalsize
	Here, the inequality \eqref{eq:ineq proof} is because, for all $y^{\prime}\in\X_H$: $y^{\prime} \neq y$, \eqref{eq:ineq max} holds and therefore $\sum_{x\in\X_H}r_{Y|X}\left(y^{\prime}|x\right)\left(P(\X_H)\PXgS(x|s)-P(\X_H)\PX(x)\right)<0$. For $s=s_u$, \eqref{eq:last proof} will be $0$, i.e., \eqref{eq:ineq max} does not hold, which is a contradiction. To show the contradiction in \eqref{eq:ineq min}, consider the numerator of \eqref{eq:17} and replace $s_u$ with $s_\ell$. We have
	$$\begin{aligned}
		&\sum_{x \in \X_H} r_{Y|X}(y|x)\left(P\left(\X_H\right)\PXgS(x|s)-P\left(\X_H|s_\ell\right) \PX(x)\right) \\
		&=\sum_{x \in \X_H}\left(1-\sum_{y^{\prime} \in \X_H: y^{\prime} \neq y} r_{Y|X}\left(y^{\prime}|x\right)\right)\\
		&\times \left(P\left(\X_H\right) \PXgS(x|s)-P\left(\X_H|s_\ell\right) \PX(x)\right) \\
		&<\sum_{x \in \X_H}\left(P\left(\X_H\right) \PXgS(x|s)-P\left(\X_H|s_\ell\right) \PX(x)\right) \\
		&=P\left(\X_H\right)\left(P\left(\X_H|s\right)-P\left(\X_H|s_\ell\right)\right)
	\end{aligned}$$
	which contradicts \eqref{eq:ineq min} for $s=s_\ell$. Therefore, Proposition \ref{prop:Xinvarinat} holds.

\bibliographystyle{IEEEtran}                
\bibliography{Asym_Watchdog,IEEEabrv}

\end{document}